\documentclass[preprint,12pt]{elsarticle}

\usepackage{graphicx}
\usepackage{dcolumn}
\usepackage{bm}
\long\def\/*#1*/{}
\usepackage[utf8]{inputenc}
\usepackage{etoolbox}
\usepackage[T1] {fontenc}
\usepackage{xcolor}
\usepackage{graphicx}
\usepackage{datetime}
\usepackage{geometry}

\usepackage{amsmath,amsfonts,amssymb}
\usepackage{fancyhdr}
\usepackage{subcaption}
\graphicspath{ {../Figures/} }
\usepackage{float}

\usepackage{mathtools}

\newcommand{\supplementarysection}{%
  \setcounter{figure}{0}
  \let\oldthefigure\thefigure
  \renewcommand{\thefigure}{S\oldthefigure}
  \section{Supplementary section}
}

\makeatletter
\def\@email#1#2{%
 \endgroup
 \patchcmd{\titleblock@produce}
  {\frontmatter@RRAPformat}
  {\frontmatter@RRAPformat{\produce@RRAP{*#1\href{mailto:#2}{#2}}}\frontmatter@RRAPformat}
  {}{}
}%
\makeatother

\begin{document}

\begin{frontmatter}

\title{Molecular Integral Equations Theory in the Near Critical Region of CO$_{2}$} 


\author[1]{Mohamed Houssein Mohamed}
\affiliation[1]{organization={Université de Lorraine and CNRS, LPCT UMR 7019}, addressline={F-54000, France}}
\author[2]{Luc Belloni}
\affiliation[2]{organization={LIONS, NIMBE, CEA, CNRS, Université Paris-Saclay}, addressline={Gif-sur-Yvette 91191, France}}
\author[3]{Daniel Borgis}
\affiliation[3]{organization={PASTEUR, Département de Chimie, École Normale Supérieure, PSL University, Sorbonne Université, CNRS}, adddressline={Paris 75005, France}}
\author[1]{Francesca Ingrosso}
\author[1]{Antoine Carof\corref{cor1}}
\ead{antoine.carof@univ-lorraine.fr.}
\cortext[cor1]{Corresponding author}
\newpageafter{author}

\begin{abstract}
Environmental concerns are driving the search for greener yet efficient solvents. Supercritical CO$_2$ (scCO$_2$) is a promising candidate due to its non-toxicity and the potential for reusing CO$_2$ emissions. It also offers a versatile range of properties that can be finely tuned by pressure adjustments. This adaptability is exploited in chemical industry processes such as separation or extraction. The development of new green processes using scCO$_2$ requires an efficient tool for predicting the solvation properties under different conditions. Existing parametric equations for solubility prediction depend on known experimental data, while molecular dynamics (MD) simulations remain expensive for studying different conditions. Both methods are unsuitable for advancing new technologies. The molecular density functional theory (MDFT) offers a promising alternative, combining an accurate microscopic modeling with ultra-fast calculations. MDFT necessitates the bulk direct correlation functions, which can be calculated from expensive MD simulations or from approximate yet rapid molecular integral equation theories. The development of MDFT as a powerful tool to study the solvation in scCO$_2$ will require the construction of an accurate molecular integral equations for scCO$_2$. In this perspective, this paper presents the exact direct correlation functions of scCO$_2$ obtained from MD and compares them with the results of the simplest molecular integral equations, the hypernetted chain approximation (HNC). If HNC fails to provide correct long-range correlations and thermodynamics, it succeeds in reproducing the short-range structure. By using the direct correlation functions obtained from MD and HNC, we demonstrate the efficacy of MDFT in calculating the chemical potential of CO$_2$ in scCO$_2$. The results open the door to the application of MDFT to a wider range of solutes dissolved in scCO$_2$, taking into account different thermodynamic conditions.  
\end{abstract}

\begin{graphicalabstract}
\includegraphics[scale=0.5]{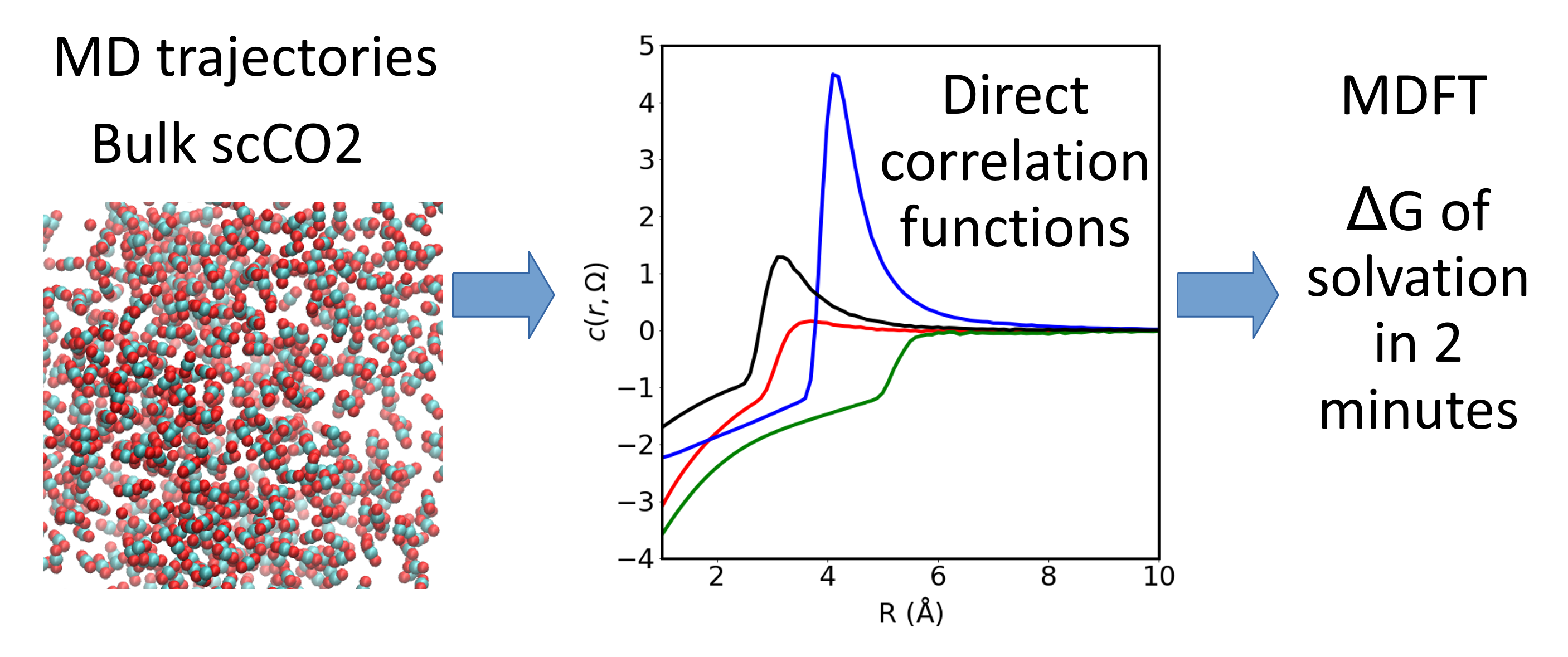}
\end{graphicalabstract}

\begin{highlights}
\item The combination of MD simulations and liquid-state theories permits to extract the correlations functions in the near-critical region of the supercritical CO$_2$
\item The molecular HNC theory reproduces well the MD-based correlation functions at all densities, except at long distances
\item In the gas-like supercritical domain, the chemical potentials of CO$_2$ calculated with the classical DFT are in excellent agreement with the MD calculation
\end{highlights}


\end{frontmatter}

\date{\today}

\newpage

\section{Introduction}
\label{sec:intro}
The properties of supercritical fluids exhibit a gradual transition from liquid-like to gas-like and are easily controlled by applying small pressure changes. The possibility of fine-tuning the density changes in the supercritical phase can be exploited to modify their dissolving power, while their gas-like behavior ensures a better mass transport (and lower viscosity) than in the liquid phase.~\cite{eckert_supercritical_1996}
These properties provide a wide range of possible applications in the chemical industry (e.g. separation and extraction processes).~\cite{kiran_supercritical_2000} In recent years, growing concerns about the environmental impact of industrial activities have drawn increasing attention on the
design of technologies following the guidelines of Green Chemistry.~\cite{anastas_green_2010}The Green Chemistry advocates for the development of a new range of solvents, that are “greener” but as efficient as the widely used organic solvents. Solvents such as CO$_2$ in the supercritical phase (scCO$_2$)
are attractive candidates because they are non-toxic to humans and the environment. In addition, scCO$_2$ is not flammable, its supercritical condition is easily accessible (the critical temperature is $\mathrm{T_{c}} = 304$~K, the critical pressure $\mathrm{P_{c}} = 73$~bar and the critical number density $n_c = 10.6$~mol.L$^{-1}$) and its use as a solvent may represent a possible reuse of the CO$_2$ emitted by industries.
Given its low solvation power, especially in the case of large and polar solutes, scCO$_2$ is often mixed with a polar co-solvent (water, acetone, methanol) at low concentrations (< 10 mol \%). These mixtures are routinely used for extraction of natural products from plants,~\cite{reverchon_supercritical_1997, herrero_sub-_2006, aydi_supercritical_2020} for the impregnation of polymers in medical applications,~\cite{champeau_drug_2015, chauvet_extrusion_2017}, and for the synthesis of nanoparticles (e.g., drug carriers).~\cite{tabernero_supercritical_2012} The promotion of new, environmentally friendly scCO$_2$-based processes requires however a molecular understanding of the solvation properties of this medium and an efficient tool for an accurate prediction of the solubilities of different species in the supercritical region.\par 

In the context of industrial applications, solubility predictions rely on parametric phenomenological equations. The solubility is expressed as a function of the pressure, the temperature, and of several parameters that must be refitted for each solute. The solubility functions can be derived from macroscopic thermodynamics,~\cite{colussi_comparison_2006, garlapati_temperature_2009, vega_perspectives_2018} from semi-empirical approaches,~\cite{chrastil_solubility_1982, kumar_modelling_1988, mendez-santiago_solubility_1999} or from simple solvation models.~\cite{shin_development_2001, cheng_calculation_2003, su_correlation_2007}  These functions accurately reproduce the experimental solubilities (with <10 \% of error on average), but they are limited to systems (solute, solvent, cosolvent) for which some experimental data are already available, which prevents these models from being used to promote the design of new technologies.  \par 

Alternatively, molecular dynamics (MD) is based on a microscopic modeling that allows accurate calculation of solvation properties and can be easily adapted to different environments (solute, co-solvent). In MD, the microscopic interactions are determined from electronic structure calculations (ab initio MD) or from force field functions (classical MD). Ab initio MD is accurate but expensive, which limits its applicability to systems with less than 300 CO$_2$ molecules.~\cite{saharay_ab_2004, saharay_evolution_2007, balasubramanian_ab_2009, mi_ab_2019}
However, the calculation of the thermodynamic properties require larger simulation boxes to converge in the near-critical region, which forces the use of classical MD. Several force fields have been developed to model the gas phase and condensed phase properties of CO$_2$.~\cite{harris_carbon_1995, potoff_vaporliquid_2001, zhang_optimized_2005, nieto-draghi_thermodynamic_2007, qin_molecular_2008, zhu_fully_2009, cygan_molecular_2012, mognetti_efficient_2008} They have been used to determine the solvation properties of various solutes (e.g., rare gases, naphthalene, anthracene, caffeine, etc.) with a reasonable accuracy (less than 2 kJ.mol$^{-1}$ for the solvation free energy).~\cite{stubbs_partial_2005, su_simulations_2006, noroozi_solvation_2016, noroozi_microscopic_2017, reddy_solubility_2019} MD simulations can also provide insights into all solvation properties: solvation enthalpy, partial molar volume, solvation structure around the solute and spectroscopic properties.~\cite{iwai_molecular_1997, stubbs_partial_2005, kajiya_investigation_2013, noroozi_microscopic_2017, altarsha_new_2012, altarsha_cavity_2012, ingrosso_driving_2016, ingrosso_modeling_2017, azofra_theoretical_2013, san-fabian_theoretical_2014, ingrosso_electronic_2018} 
Despite the increasingly availability of computational resources, the calculation of solvation properties with MD remains expensive, requiring several tens hours.CPU. If this moderate cost allows for a detailed study of solvation in scCO$_2$ for a given solute, it cannot be afforded for the investigation of a large number of solutes under different conditions (pressure, temperature, fraction of co-solvent).  \par

Liquid-state theories offer the best of both worlds. These powerful modeling techniques rely on a direct integration of the liquid partition function with a microscopic Hamiltonian (usually described with the same force fields as in MD).~\cite{hansen_theory_2013} 
This strategy allows to avoid the computation of the full microscopic trajectory, leading to much faster calculations than MD (at least 1000 times faster),~\cite{hansen_theory_2013} while maintaining the same level of microscopic description and the same flexibility to explore complex systems as MD. However, the direct integration cannot be performed analytically and many different approximations and mathematical frameworks have been developed in the literature to approximate the exact results. These approximations can be divided into bulk theories, which aim to calculate 2-body correlation functions in an homogeneous liquid or mixture,~\cite{hansen_theory_2013} and classical density functional theory (cDFT), which aims to determine the 1-body density in the presence of an external potential.~\cite{evans_nature_1979} Among these methods, cDFT is a suitable framework to investigate solvation. cDFT requires two inputs: the solute-solvent interactions, exactly known, and the solvent-solvent effective interactions, which are unknown. With the appropriate level of approximation for the latter, the cDFT will constitute the optimal balance between efficiency and accuracy for the development of original tools and their applications to solvation properties in scCO$_2$.  \par

In previous studies, cDFT has been applied to investigate supercritical fluids with two crucial approximations: (i) the solvent is spherical and (ii) solvent-solvent interactions are treated with a perturbative approach.~\cite{hansen_theory_2013}  The latter approach takes the hard-sphere fluid as a reference, since accurate density functionals exist for this fluid,~\cite{rosenfeld_free-energy_1989, roth_fundamental_2010} while attractive interactions are included with a mean-field description. Egorov employed this strategy to explore the original characteristics of the solvation in supercritical fluids, such as the local density augmentation near the solute.~\cite{egorov_local_2000, egorov_preferential_2000} Mognetti et al. developed a spherical force field for scCO$_2$ by averaging over the angles the quadrupolar interactions between CO$_2$ molecules.~\cite{mognetti_efficient_2008} With this coarse-grained force field, they construct an integral equation theory to reproduce the thermodynamics of scCO$_2$,~\cite{mognetti_spherically_2008} but do not apply it to study solvation processes. More recently, Budkov et al. used a spherical description and perturbative strategy to determine the qualitative evolution of the solvation free energy in supercritical CO$_2$ for different pressures and temperatures.~\cite{budkov_possibility_2019, kalikin_carbamazepine_2020, kalikin_computation_2021} They focus in particular on molecules of pharmaceutical interest (aspirin, ibuprofen, carbamazepine).~\cite{budkov_possibility_2019, kalikin_carbamazepine_2020, kalikin_computation_2021} However, considering CO$_2$ as a spherical particle implies to neglect the orientational interactions among CO$_2$ molecules and the solute, which prevents the accurate description of specific interactions. This is particularly relevant, given the importance of specific solute-solvent interactions in promoting a better solubilization in the CO$_2$ medium, as it is done by using CO$_2$-phile solutes.~\cite{ingrosso_modeling_2017, lee_combined_2015, girard_structureproperty_2016}  \par


Two frameworks exist in liquid theories to overcome the spherical description of molecules: the reference interaction-site model (RISM) and the molecular density approach. RISM extends various bulk integral equation theories for multi-site systems and is widely used to study solvation in water.~\cite{chandler_optimized_1972, hirata_extended_1981, beglov_integral_1997, kovalenko_self-consistent_1999, ratkova_solvation_2015, giambasu_predicting_2019} The RISM equations lack of consistency and ad hoc corrections must be added. On the other hand, liquid-state theories can be written in function of the molecular density $\rho(\mathbf{r}, \omega)$, which depends on both the molecular positions $\mathbf{r}$ and on the molecular orientations $\omega$. The molecular density $\rho(\mathbf{r}, \omega)$ naturally incorporates the molecular structure. Some of us have recently developed a cDFT for the molecular density, called molecular DFT (MDFT), and they described the solvent-solvent interactions within the homogeneous reference approximation (HRA). The latter necessitates only the direct pair (2-body) correlation functions (DCF) of the pure solvent,~\cite{levesque_solvation_2012} which directly relates to the pair distribution functions (PDF).The latter can be calculated from MD simulations of the bulk solvent, at a given pressure and temperature.~\cite{hansen_theory_2013, puibasset_bridge_2012, belloni_exact_2017} MDFT-HRA was applied to water, which led to qualitative successes.~\cite{jeanmairet_molecular_2013-1, levesque_solvation_2012, luukkonen_hydration_2020} Work is currently ongoing to overcome the HRA and provides more accurate approximations.~\cite{borgis_simple_2020}   \par

Based on these results, we decided to evaluate the potential of MDFT-HRA for modeling the solvation properties in scCO$_2$, and in particular their evolution with the pressure and the temperature. MDFT-HRA requires two inputs, the bulk 2-body DCFs and the CO$_2$-solute interactions. In the previous application of MDFT for water, the DCFs were extracted from bulk MD simulations at a given pressure and temperature. As our objective is to investigate the solvation properties as a function of pressure and temperature, this approach is not viable. Indeed, running a bulk MD simulation for each temperature and each pressure will result in a loss of the benefits of MDFT-HRA compared to directly simulating the solute and the solvent with MD. Conversely, an appropriate integral equation theory for bulk scCO$_2$ could provide accurate DCFs in a highly efficient manner, thus enabling to perform MDFT calculations at any pressure and temperature. Overall, we aim to design an innovative modeling of solvation in scCO$_2$ by integrating two liquid-state theories: a bulk 2-body theory for the fast determination of the DCFs, which will input MDFT-HRA for efficient and accurate calculation of the solvation properties.  \par

This paper lays the first stone of such a bulk 2-body molecular integral equation theory for scCO$_2$. It provides: (i) a set of MD-based DCFs to serve as a reference for evaluating the accuracy of the future theory, (ii) a comparison with the the hyper-netted chain approximation (HNC), an approximate but robust integral equation; and (iii) an evaluation of the performance of MDFT-HRA in reproducing the chemical potential of CO$_2$. Given that industrial applications of scCO$_2$ occur between $P_c$ and $2~P_c$, and $T_c$ and $1.2~T_c$, we decided to analyze four different densities $n_b$, $n_b = 0.4, 0.7, 0.8$ and $2.0~n_c$, with $n_c$ the critical number density, and along the isotherm $T = 1.05~T_c$. For each density, we computed the thermodynamic properties and the DCFs obtained from MD and we compared them with those predicted by HNC. This comparison assesses the limitations of HNC and pave the way for the construction of a more accurate integral equation theory. This work is concluded by testing the aforementioned strategy for the calculation of the solvation free energy of CO$_2$ in scCO$_2$ (\textit{i.e.}, the chemical potential of CO$_2$). We compared the solvation free energy calculated with three methods: MD calculations (the reference data), MDFT-HRA using the exact correlation functions and MDFT-HRA using the HNC correlation functions. The MDFT-HRA solvation free energy are in excellent agreement with the MD ones in the gas-like domain.   \par

The paper is organized as follows: in Section~\ref{sec:theory}, we will introduce the MDFT and the scheme to extract the direct correlation functions (DCFs) for a molecular liquid by combining integral equations, molecular simulations and the rotational invariant formalism. We will describe in detail the parameters of the MD simulations and of the DCFs calculation in Section~\ref{sec:simdetails}. In Section~\ref{sec:results}, we will discuss the correlation functions at different densities and compare them with the HNC prediction. We will finally conclude the work by presenting an evolution of the MDFT-HRA ability to calculate the solvation free energy of CO$_2$ in scCO$_2$. 

 \section{Theoretical Background}
\label{sec:theory}

\subsection{Molecular density functional theory}

In MDFT, the solvation free energy $\Delta G_\mathrm{solv}$ and the solvation structure $\rho_\mathrm{eq}$ are obtained by minimizing the functional $\mathcal{F}[\rho]$, 

\begin{align}
\Delta G_\mathrm{solv} = \Omega - \Omega_b = \min\{ \mathcal{F}[\rho] \} = \mathcal{F}[\rho_\mathrm{eq}]
\end{align}

\noindent where $\Omega_b$ is the grand potential of the bulk solvent and $\rho(\mathbf{r}, \Omega)$ denotes the molecular solvent density, a function of the position vector of the solvent molecules $\mathbf{r}$ and of the three Euler angles $\Omega = (\theta, \phi, \psi)$, which characterize the orientation of a solvent molecule. The MDFT functional is decomposed in three parts, 

\begin{align}
\mathcal{F} = \mathcal{F}_\mathrm{id} + \mathcal{F}_\mathrm{ext} + \mathcal{F}_\mathrm{exc}
\end{align}

\noindent
The ideal term, $\mathcal{F}_\mathrm{id}$, represents the entropy of fluid of non-interacting molecules and reads

\begin{equation}
\mathcal{F}_\mathrm{id} = k_BT \int \mathrm{d}1 \left[ \rho(1)\ln\left(\frac{\rho(1)}{\rho_b}\right) - \Delta \rho(1) \right]
\end{equation}

\noindent where $k_BT$ is the thermal energy, $\rho_b = n_b / (8\pi^2)$ is the homogeneous solvent density ($n_b$ is the number density and $8\pi^2$ is the angular normalization constant), $\Delta \rho(1) = \rho(1) - \rho_b$, and the notation ``1'' stands for the position and Euler angle coordinate of a given molecule ($1 = (\mathbf{r}, \Omega)$). The external term, $\mathcal{F}_\mathrm{ext}$, accounts for the interaction between the solvent and the solute

\begin{align}
\mathcal{F}_\mathrm{ext} = \int\mathrm{d}1 \rho(1)v_\mathrm{ext}(1)
\end{align}

\noindent where $v_\mathrm{ext}$ is the potential of the solute-solvent interactions, typically obtained from standard force fields describing the solute-solvent interactions. Finally, the excess term, $\mathcal{F}_\mathrm{exc}$, accounts for the solvent-solvent interactions.This term is the most challenging to define, as the functional form is actually unknown and must be approximated. We have followed the approach of the homogeneous reference approximation (HRA).~\cite{van_leeuwen_new_1959} This begins with a Taylor expansion with respect to the homogeneous bulk density $\rho_b$,

\begin{align}
\mathcal{F}_\mathrm{exc} &= -\frac{k_BT}{2} \int\mathrm{d}1\mathrm{d}2 \Delta\rho(1) c(12) \Delta\rho(2) + \mathcal{F}_\mathrm{B} \nonumber \\
&= \mathcal{F}_\mathrm{HRA} + \mathcal{F}_\mathrm{B} 
\end{align}

\noindent 
where the bridge functional $\mathcal{F}_\mathrm{B}$ incorporates all the terms beyond the quadratic order, while $c(12)$ are the direct correlation functions (DCFs) of the bulk solvent (without solute). HRA consists in neglecting the bridge function ($\mathcal{F}_\mathrm{B} = 0$), resulting in a MDFT functional that requires only two inputs: the external potential $v_\mathrm{ext}$ and the DCFs $c(12)$. Finally, the functional of the MDFT-HRA approach reads, 

\begin{equation}
    \mathcal{F} = \mathcal{F}_\mathrm{id} + \int\mathrm{d}1 \rho(1)v_\mathrm{ext}(1) -\frac{k_BT}{2} \int\mathrm{d}1\mathrm{d}2 \Delta\rho(1) c(12) \Delta\rho(2)
    \label{eq:mdfthra}
\end{equation}

\subsection{Molecular correlation functions}

The calculation of the bulk DCFs can be performed using either MD simulation or integral equation theories.In both cases, the Ornstein-Zernike equation~\cite{hansen_theory_2013} is employed, which establishes a link between the DCF $c$ and the total correlation function $h$, 
\begin{equation}
h(12)-c(12) = c(12) + \rho_b \int h(13)c(32) \mathrm{d}3,
\label{eq:oz}
\end{equation}

\noindent
where 1, 2 and 3 represent the position and the orientation of molecules 1, 2 and 3, respectively. The total correlation function $h$ is directly related to the pair correlation function $g$, $h=g-1$. \par

MD simulations provide an accurate sampling of the equilibrium conformations of the liquid for a given pair potential $v$,  from which the PDFs $g$ can be extracted. Assuming that the finite-size effects and the statistical noise are small enough, the so determined $g$ is exact and the OZ equation (Eq.~\ref{eq:oz}) can provide the exact DCFs $c$. On the other hand, integral equations necessitate a second equation to ``close'' the OZ (Eq.~\ref{eq:oz}),
 
\begin{eqnarray}
g(12) &=& \exp(-\beta v(12) + \ln y(12) ) \\
&=& \exp(-\beta v(12) + \gamma(12) + b(12) )
\nonumber
\label{eq:bridge}
\end{eqnarray}

\noindent 
where  $b$ is called the bridge function, $\ln y(12)$ the cavity potential and $\gamma(12)$ is the indirect correlation function. The latter can be related to the pair distribution function through the OZ convolution product (Eq.\ref{eq:oz}). 
The expression of the bridge function $b$ is usually unknown, despite the existence of some formal diagrammatic or functional exact expressions.~\cite{hansen_theory_2013} The goal of integral equation theory is to make the best possible assumption for the bridge function $b$ and to solve the corresponding approximate closure equation (Eq.~\ref{eq:bridge}) coupled with the OZ equation (Eq.~\ref{eq:oz}) through an iterative procedure. 
The simplest choice is to neglect the bridge function, $b=0$, and is called the HNC approximation. This is equivalent to the homogeneous reference approximation (HRA), but in a different framework (integral equations for HNC, cDFT for HRA). \par

Given that the CO$_2$ molecule is anisotropic, the molecular correlations $g$, $c$, and $b$ depend on the separation distance $r=r_{12}$ between particles 1 and 2, as well as on their relative orientation $\Omega$. Instead of manipulating the complete $g(r,\Omega)$ as an explicit function of the different Euler angles, it is more fruitful to project it onto a basis of angular functions and to develop new expressions in terms of projections depending exclusively on the distance $r$. Following Blum's notation and normalization~\cite{blum_invariant_1972, blum_invariant_1972-1}, we can write:

\begin{eqnarray}
g(r,\Omega) &=& g(r, \Omega_1, \Omega_2, \hat{r}) \\
&=& \sum_{m=0}^{\infty}\sum_{n=0}^{\infty}\sum_{l=\vert m-n\vert}^{m+n} \sum_{\mu=-m}^{m} \sum_{\nu=-n}^{n} g_{\mu\nu}^{mnl}(r)\Phi_{\mu\nu}^{mnl}(\Omega_1, \Omega_2, \hat{r})
\label{eq:g_projections}
\nonumber
\end{eqnarray}

\noindent
with

\begin{eqnarray}
\Phi_{\mu\nu}^{mnl}(\Omega_1, \Omega_2, \hat{r}) = \sqrt{(2m+1)(2n+1)}\sum_{\mu', \nu', \lambda'} \begin{pmatrix} m & n & l  \\
  \mu' & \nu' & \lambda'  \end{pmatrix} R_{\mu'\mu}^m(\Omega_1)R_{\nu'\nu}^n(\Omega_2)R_{\lambda'0}^l(\hat{r})  
\label{eq:rot_invariant}
\end{eqnarray}

and similar expansions apply to any correlation function. The coefficients $\begin{pmatrix} m & n & l  \\
\mu' & \nu' & \lambda'  \end{pmatrix}$ are the 3-j symbols, and $R_{\mu'\mu}^m(\Omega)$ are Wigner generalized spherical harmonics (definition and notation from Messiah~\cite{messiah_quantum_1962}). The functions $\Phi_{\mu\nu}^{mnl}$ are the rotational invariants (independent of the reference frame) and they form an orthogonal basis.~\cite{blum_invariant_1972, blum_invariant_1972-1} They depend on the relative orientation of the two molecules and of the vector connecting them (five Euler angles) and are characterized by five indices $m, n, l, \mu, \nu$. For linear particles with a center of symmetry, the following conditions apply: the projections are real, $m+n+l$  is even, $m$ and $n$ are even and $\mu=\nu=0$. In the following, the indices $\mu$, $\nu$ are omitted for clarity. The coefficients $g^{mnl}(r)$ are derived from the complete function by angular projection:

\begin{equation}
g^{mnl}(r) = \frac{\int\int\int g(r,\Omega)\Phi^{mnl*}(\Omega)\mathrm{d}\Omega}{\int\int\int \vert \Phi^{mnl}\vert^2(\Omega)\mathrm{d}\Omega} = (2l+1) \int\int\int g(r,\Omega)\Phi^{mnl*}(\Omega)\mathrm{d}\Omega.
\label{eq:gmnl}
\end{equation}

\noindent
Using a more compact notation form now on for CO$_2$ ($\Phi^\alpha$ stands for $\Phi^{mnl}$), the expansion and the projections read:

\begin{eqnarray}
g(r, \Omega) &=& \sum_{\alpha}g^\alpha(r) \Phi^\alpha(\Omega),  \\
g^\alpha &=& \left\langle g(r, \Omega) {\Phi^*}^\alpha(\Omega) \right\rangle,
\label{eq:galpha}
\end{eqnarray}

\noindent
where the brackets in Eq.~\ref{eq:galpha} represent a (normalized) triple angular integral. The first projection $g^{000}$ represents the center of mass-center of mass PDF (averaged over all orientations). 

In principle, the expansion in Eq.\ref{eq:galpha} runs over an infinite basis set, but in practice is limited to a finite number $\alpha_\mathrm{max}$ of projections, characterized by $m, n \le n_\mathrm{max}$. For the present CO$_2$ case, we found that $n_\mathrm{max}=6$ (\textit{i.e.} $\alpha_\mathrm{max}=30$ independent projections) is a reasonable choice (results not shown) for capturing the angular dependence of the correlations, at least for the excess potential of mean force $\ln~y$, see below for more details.

\section{Numerical Procedure}
\label{sec:simdetails}

\subsection{Pair potential and thermodynamic condition}
\label{sec:potential}

In this work, we have studied four densities in the near-critical conditions ($0.4~n_{c}$, $0.6~n_{c}$, $0.8~n_{c}$ and $2~n_{c}$) at $T = 1.05~T_c$. Among the force fields for CO$_2$ available in the literature~\cite{harris_carbon_1995, potoff_vaporliquid_2001, zhang_optimized_2005, cygan_molecular_2012}, we chose the rigid version of the EPM2 model by Harris and Yung,~\cite{harris_carbon_1995} whose parameters are reported as supporting information (Table~S1). This model  accurately reproduces the phase diagram of CO$_2$ in the near-critical region with a simple description of the molecular interactions. It describes CO$_2$ as a linear molecule with fixed C-O distances, $d_{\mathrm{CO}}= 1.149$~\AA.  The intermolecular interactions are obtained as the sum of the Lennard-Jones and Coulomb pair potentials. The critical values for this force field are: $T^\mathrm{EPM2}_c = 313.4$ K, $n_c^\mathrm{EPM2} = 10.3$ mol.L$^{-1}$, and $P_c^\mathrm{EPM2} = 76.5$ bar.  \par

\subsection{Molecular dynamics simulations}
\label{sec:md_details}

For each density, the cubic simulation boxes contained 4000 CO$_2$ molecules, except for $n_b = 0.8~n_{c}$, for which a larger number of molecules (13500) was needed to capture the near-critical fluctuations. The box lengths $L_\mathrm{box}$ are $116.0$, $96.3$, $138.2$ and $67.8$~\AA~ for $n_b =$ $0.4$, $0.7$, $0.8$ and $2.0~n_c$, respectively. Periodic boundary conditions were applied in all directions. The long-range electrostatic interactions were treated using the Ewald summation with a particle-particle particle-mesh solver, with a cutoff of $r_\mathrm{cut} = 17~$\AA~and an accuracy of $\mathrm{10}^{-5}$.~\cite{hockney_computer_1988} A cutoff of the direct long-range LJ interactions beyond $r_\mathrm{cut} = 17~$\AA~was introduced, and mean-field corrections for the energy and the pressure were added .~\cite{sun_compass_1998}  \par

We performed the simulations with the LAMMPS software.~\cite{thompson_lammps_2022} After an equilibration step of 2 ns (NVT ensemble with $T = 320~K$), the trajectories were propagated for 6 ns (NVT ensemble with $T = 320~K$) and the atomic positions were printed every 5~ps, resulting in 1200 independent configurations. The equilibrium pressures are 76, 103, 97 and 442~bar for the densities $n_b =$ $0.4$, $0.7$, $0.8$ and $2.0~n_c$, respectively. The equations of motion were integrated using the velocity Verlet algorithm with a timestep of 1~fs and a Nosé-Hoover-like thermostat.~\cite{parrinello_polymorphic_1981, martyna_constant_1994, shinoda_rapid_2004} The same protocol applies to all densities. \par

Along each MD trajectory, the desired projections of the PDFs are constructed from the relative positions and orientations of the particle pairs in the box. For efficiency, we first compute the projections in each intermolecular frame, the so-called intermediate projections $g_\chi^{mn}(r)$.~\cite{blum_invariant_1972-1} To do this, the minimum image distances $r_{ij}$ between carbon atoms of all pairs $i<j$ of molecules are sorted into a histogram of width $\delta r=0.025$~\AA. The bin $k$ is defined as the distance interval $I_k=\left[r_{k-1},r_k\right]$, with $r_k=k \delta r$, corresponding to the volume of the spherical cap $V_k=4\pi/3(r_k^3-r_{k-1}^3)$. The intermediate projections (in the intermolecular frame) are constructed according to:

\begin{equation}
    g_\chi^{mn}(\mathrm{bin}~k) = \frac{V}{1/2N^2 V_k} {\sum_{i<j}}_{r_{ij} \in I_k} f_m f_n {R^m_{\chi 0}}^*(\Omega_i') {R^n_{-\chi 0}}^*(\Omega_j')
\label{eq:gbink}
\end{equation}

\noindent
where $\Omega'$ is the orientation of the molecule in the local frame and the $f_m$ are the Blum's normalization factors, $f_m = (2m+1)^{1/2}$. The spherical harmonics ${R^m_{\chi O}}(\Omega_i')$ are calculated recursively from $m=1$ to $m=n_\mathrm{max}$, without any trigonometric manipulation.~\cite{choi_rapid_1999} The bin $k$ is then assigned to the distance $r=r_k-\delta r/2$. In a second step, the $g^{mnl}(r)$ projections in the laboratory frame are recovered from the $g^{mn}_{\chi}(r)$ intermediate projections using the $\chi$-transform as defined by Blum.~\cite{blum_invariant_1972-1} Since the MD is performed under NVT conditions, the derived PDFs suffer from explicit finite-size errors  due to the absence of fluctuations in the density.~\cite{belloni_finite-size_2017} The main effect concerns the long-range asymptote of $g^{000}$, which differs from the theoretical value of 1, and is instead, 

\begin{equation}
    \lim_{r \rightarrow \infty} g_N^{000}(r) = 1 - \frac{\chi_T}{N}
\label{eq:glimit}
\end{equation}

\noindent
where $\chi_T = \left( \frac{\partial \rho}{\partial \beta P} \right)_T $ is the normalized isothermal compressibility which is related to the center of mass structure factor $S^{000}(q)$ at zero $q$, $\chi_T = S^{000}(0) = 1 + \rho_b \int_0^\infty(g^{000}-1)4\pi r^2 \mathrm{d}r$. For the forthcoming use of the MD data within the integral equation machinery, it is important to correct for this bias. A simple procedure consists in dividing the measured PDFs $g$ by the factor $1-\chi/N$. Since the value of $\chi$ is unknown a priori, one can either extract its value from a careful estimate of the asymptote, when a clear long-range plateau is observed in the case of a sufficiently large simulation box, or compute it in self-consistent manner from the running integral of $g-1$. Note that, for conditions close to the critical point, the correlations become long-ranged and the values of the compressibility can be high. Hence, the correction is not negligible, even for large boxes containing tens of thousands of particles.  \par

The next step to determine the DCFs is to insert the exact MD-based PDFs $g$ into the OZ equation (Eq.~\ref{eq:oz}). However, the inversion of the OZ relation from range-limited MD data is not trivial. We overcome this difficulty by combining the MD data with integral equation theory. Before detailing this procedure, we will first explain the resolution of the standard HNC equations. 

\subsection{Solution of the HNC equations}
\label{sec:hnc_resol}

The iterative resolution of Eq.~\ref{eq:oz} with the closure relation (Eq.~\ref{eq:bridge}) and the HNC assumption $b=0$ is a standard technique, although it has revealed quite expensive for anisotropic interactions.~\cite{patey_integral_1977, lado_integral_1982, fries_solution_1985, fries_resolution_1987, anta_fast_1995, lado_integral_1995} One of us recently proposed a powerful algorithm to speed up the iterative resolution of anisotropic HNC.~\cite{belloni_efficient_2014} A cycle starts with a $\gamma^\mathrm{in}$ estimate of the $\gamma$ projections in the $r$ space.  The closure (Eq.~\ref{eq:bridge}) produces the $g$ projections through a transitory exchange with the angle description. The $c=g-1-\gamma$ projections are then numerically Fourier-Hankel transformed. In the following, we will denote a Fourier-transformed function with a hat and note the wavenumber as $\mathbf{q}$. In the Fourier space and the intermolecular frame, the OZ equation (Eq.\ref{eq:oz}) consists of algebraic products between projections and is easily solved by matrix inversions. The $\hat{\gamma}$ function is then inversely Fourier-Hankel transformed to produce $\gamma^\mathrm{out}$ projections. If the "in" and "out" sets differ by more than a value fixed through a convergence criterion (e.g., $\gamma^\mathrm{out} - \gamma^\mathrm{in} < 10^{-8}$), a new cycle is started. Each cycle consists of one Picard procedure (with the mix parameter 0.6) and several (usually 5) Zerah linear cycles (parameter: 0.001).~\cite{belloni_efficient_2014} The HNC algorithm typically converges in 3-5 cycles and takes about one minute.  \par

The main difficulty is to find a first input function $\gamma^\mathrm{in}$ to ensure the convergence of the procedure. In the gas-like region of the isotherm (below $0.8~n_c$), we have adopted the following strategy: we first solved the HNC equations for a low density ($0.1~n_c$), using the ideal gas result as input ($\gamma^\mathrm{in} = 0$). After the successful convergence of the HNC algorithm, we obtained the $\gamma$ function at this density. We then increased the density step by step ($n_i = n_{i-1} + \Delta n$, $\Delta n = 0.1$), and used the converged $\gamma$ function of the density $n_{i-1}$ as an input for the new density $n_i$. This method stops at $0.7~n_c$, beyond which one enters the forbidden HNC two-phase region. In the liquid-like region, we started from a high density $2.0~n_c$, far from the HNC-forbidden domain. However, at this density, the ideal gas-phase $\gamma$ fails as an input. We got around this problem by first solving HNC with a ``faded'' interaction (${\epsilon'}{_\mathrm{LJ}}= \epsilon_\mathrm{LJ}/2$ and ${\sigma'}{_\mathrm{LJ}} = \sigma_\mathrm{LJ}/2$), for which HNC converges with the ideal gas input ($\gamma^{in} = 0$). Again, we followed an incremental strategy: using the previously converged $\gamma$ as an input, we then gradually increased $\epsilon_{LJ}$ and $\sigma_{LJ}$, solved HNC again, and continued until ${\epsilon'}_\mathrm{LJ}$ and ${\sigma'}_\mathrm{LJ}$ reached the values of the EPM2 force field. Finally, from the converged $\gamma$ at $2.0~n_c$ with the full potential, we gradually decreased the density down to $0.9~n_c$, below which no HNC solutions exist.

\subsection{Ornstein-Zernike inversion with MD pair distribution functions}
\label{sec:oz}

Let us now come back to the calculation of the exact DCFs, using the MD-calculated PDFs (see Sec.~\ref{sec:md_details}) and the OZ equation (Eq.~\ref{eq:oz}). Since the simulation data are available in a limited range of distances $r$, i.e. $r<L_\mathrm{box}/2$, the long-range tail of the PDFs cannot be computed, and their Fourier-Hankel transform leads to unwanted and uncontrolled oscillations at low wavenumbers $q$, thus preventing the resolution of the integral equation. By exploiting the \textit{a priori} exact statement that $b$ is shorter ranged than $g$, we circumvented this problem by using the MD data available at short-distance ($r<r_\mathrm{max}$) and neglecting $b$ at longer distances ($r>r_\mathrm{max}$), thus defining the new mixed closure:

\begin{eqnarray}
g^\alpha(r) &=& g^\alpha(r) (\mathrm{MD}), \;\;  \mathrm{r \leq r_\mathrm{max}}  \\
&=& \left\langle \exp\left[ - \beta v(r, \Omega) + \sum_{\alpha'}\gamma_{\alpha'}(r)\Phi_{\alpha'}(\Omega) \right] \Phi_\alpha^*(\Omega) \right\rangle, \;\;  \mathrm{r > r_\mathrm{max}}
\label{eq:mixed_closure}
\end{eqnarray}

\noindent
where $r_\mathrm{max} < L_\mathrm{box}/2$. The iterative solution with respect to $\gamma(r > r_\mathrm{max})$ is similar to that described in section~\ref{sec:hnc_resol}. The absence of any discontinuity in the $g^\alpha(r)$ functions at $r_\mathrm{max}$ \textit{a posteriori} will confirm the neglect of the bridge beyond $r_\mathrm{max}$. Note that introducing the maximum of available MD data, by choosing $r_\mathrm{max}$ close to $L_\mathrm{box}/2$, may be counterproductive, because the long-range $g^{MD}(r)$ is perturbed by the environment around the images of the neighboring boxes (implicit finite-size effects). In practice, we choose $r_\mathrm{max} = 30$, 30, 45 and 20 \AA~for the densities 0.4, 0.7, 0.8, and 2.0~$n_c$, respectively. The resolution of the molecular OZ with MD data will provide the exact DCFs $c_\alpha$, which are the main ingredient for the MDFT-HRA functional (Eq.~\ref{eq:mdfthra}). \par

\subsection{Extraction of $\ln~y$ at short and long distances}
\label{sec:extraction_short}

In addition to the DCFs $c_\alpha$, we are also interested in measuring the exact bridge $b(r<r_\mathrm{max})$ and the correlation functions for different orientations $\Omega$. In both cases, the projections of $\ln~y$ are required. However, their calculation cannot be carried out using the direct method (collecting the projections of $g$ to form $g(r,\omega)$, taking the logarithm, adding the potential and projecting the difference). Indeed, the chosen truncated basis of projections {$n_\mathrm{max}$} is not large enough for the $g$ angular description, which is very anisotropic due to the $v$ contribution and the exponentiation (the basis set is large enough for the convolutions $\gamma$, $\ln y$ and $b$). The extractions of the $\ln y$ projections from the MD-computed $g$ are achieved by solving the following implicit equations at each distance $r$:

\begin{eqnarray}
    g^\alpha &=& \left\langle \exp\left[ - \beta v(r, \Omega) + \sum_{\alpha'}[\ln~y]_{\alpha'}(r)\Phi_{\alpha'}(\Omega) \right] \Phi_\alpha^*(\Omega) \right\rangle = g^\alpha(\mathrm{MD})
\label{eq:yimplicit}
\end{eqnarray}

The fast iterative solution of such equations benefits from powerful Newton-Raphson techniques. Once the convergence is reached and the $[\ln~y]_\alpha$ obtained, the projections of $b$ are obtained by difference, $b_\alpha=[\ln~y]_\alpha-\gamma_\alpha$.~\cite{belloni_efficient_2014} The procedure breaks down however for small values of $r$, when two CO$_2$ molecules overlap for some orientations, for which $v(\Omega)\approx + \infty$ and $g^\mathrm{MD}(\Omega)=0$. In practice, the projections of $\ln y$ and $b$ are only available at $r>4.2$~\AA~ and short distances are missing. \par


To complete the $[\ln~y]_\alpha$ at short distances, we measured the cavity functions $y(r,\Omega) = \exp(\ln~y)$ during the simulation. Unlike the PDFs, these functions are not affected by the Boltzmann factor $\exp(-\beta v)$ and the overlap of CO$_2$ molecules. The  Henderson technique is the method of choice to calculate $y(r,\Omega)$ at lower density conditions.~\cite{henderson_test_1983} We reproduce here the procedure previously applied for dipolar fluids.~\cite{puibasset_bridge_2012} A test CO$_2$ molecule "0" is randomly inserted (in position and orientation) every 5~ps inside the configuration of the MD trajectory. The interaction energy $v_\mathrm{test}$ felt by the test molecule 0 and exerted by the ensemble of particles in the bulk fluid is calculated. According to the Widom insertion technique,~\cite{widom_topics_1963} the average $\left\langle \exp(-\beta v_\mathrm{ext}) \right\rangle$ gives the inverse activity coefficient $\exp(- \beta \mu_\mathrm{exc})$ with $\mu_\mathrm{exc}$ the excess chemical potential. The short-range PDFs $g(r,\Omega)$ can be derived from the relative position/orientation $r=r_{0i}$, $\Omega=\Omega_0$, $\Omega_i$,  between the test molecule 0 and each molecule $i$ of the fluid, following $g(r, \Omega)=\left\langle \exp(-\beta v_\mathrm{ext}(r, \Omega) \right\rangle / \left\langle \exp(-\beta v_\mathrm{ext}) \right\rangle $. In practice, the (intermediate) projections $g^\alpha$ are first constructed in the intermolecular frame and for each molecule $i$ of the fluid, as

\begin{equation}
g_\chi^{mn}(\mathrm{bin} k) = \frac{V}{N V_k \left\langle \exp(-\beta v_\mathrm{ext}) \right\rangle} {\sum_i}_{r_{0i} \in I_k} \exp\left[-\beta v_\mathrm{test}(\Omega_0', \Omega_i')\right] f_m f_n {R^m_{\chi 0}}^*(\Omega_0') {R^n_{-\chi 0}}^*(\Omega_i')
\label{eq:gbink2}
\end{equation}

For the projections $y_\alpha$, it is sufficient not to count the contribution of the particle $i$ to the interaction felt by the test particle 0, by defining $v_\mathrm{test}^{*i}$, with $v_\mathrm{test} = v_\mathrm{test}^{*i} + v_{0i}$:

\begin{equation}
y_\chi^{mn}(\mathrm{bin} k) = \frac{V}{N V_k \left\langle \exp(-\beta v_\mathrm{ext}) \right\rangle} {\sum_i}_{r_{0i} \in I_k} \exp\left[-\beta v_\mathrm{test}^{*}(\Omega_0', \Omega_i')\right] f_m f_n {R^m_{\chi 0}}^*(\Omega_0') {R^n_{-\chi 0}}^*(\Omega_i')
\label{eq:gammabink}
\end{equation}

Concretely, one performs a few thousand (independent) random insertions per retained bulk configuration. It may happen that the test particle 0 is inserted close to a particle $i$, resulting in $\beta v_\mathrm{test} \gg 1$ and $\exp(-\beta v_\mathrm{test}) = 0$. Such situation corresponds to an overlap between the test and the particle $i$ (in practice, we define an overlap with the criteria $\beta v_\mathrm{test} > C$, with $C = 20$). If the test particle 0 produces an overlap with at least two particles simultaneously, no information is gained for $\mu^\mathrm{exc}$, $g$ and $y$. On the other hand, if there is a single overlap say with particle $i$, information is gained only for $y$, from the $0i$ pair only. Finally, if there is no overlap at all, the statistics for the three functions are improved, from all $0i$ pairs. Typically, at $n_b =0.7~n_c$, these three situations occur in 10~\%, 40~\%, and 50~\% of the insertion attempts, respectively. Again, the Henderson technique is efficient here because the CO$_2$ fluid under study is not very dense. In comparison, it is useless for the case of bulk water at room temperature and pressure where a completely different technique is required.~\cite{belloni_exact_2017}

At the end of the Henderson procedure, the Henderson $g^\alpha(r)$ projections should coincide with those measured by the standard histogram technique (detailed in Sec.~\ref{sec:md_details}). Finally, the $\ln y_\alpha(r)$ projections are extracted from the Henderson $y_\alpha(r)$ set by following an implicit equation, similar to Eq.~\ref{eq:yimplicit}:

\begin{eqnarray}
    y_\alpha &=& \left\langle \exp\left[ \sum_{\alpha'=1}^{\alpha_\mathrm{max}}\ln y_{\alpha'}(r)\Phi_{\alpha'}(\Omega) \right] \Phi_\alpha^*(\Omega) \right\rangle \\
    &=& y_\alpha(\mathrm{MD})
    \nonumber
\label{eq:yalpha}
\end{eqnarray}

This time, the entire $r$ range is reached, including the overlapping short-range region.

\subsection{Calculation of orientational correlation functions}

The calculation of the orientational PDFs $g(r, \Omega)$ cannot be done directly by summing the $g^\alpha$, because the truncated basis of projections {$n_\mathrm{max}$} is too small for the convergence of Eq.~\ref{eq:g_projections}. On the other hand, the sum of $[\ln~y]_\alpha$ converges (Eq.~\ref{eq:g_projections}) to give the cavity potential, $\ln~y(r, \Omega) = \sum_\alpha [\ln~y]_\alpha(r) \Phi^\alpha(\Omega)$. The implicit equation for $\ln~y$ and the Henderson technique explained in Sec.~\ref{sec:extraction_short} allow to calculate $[\ln~y]_\alpha$. This permits to evaluate the orientational PDFs as,

\begin{eqnarray}
g(r, \Omega) = \exp\left( - \beta v(r, \Omega) + \ln~y(r, \Omega) \right) 
\label{eq:gromega}
\end{eqnarray}

Similarly, the orientational DCFs $c(r, \Omega)$ cannot be obtained from the $c_\alpha$, but from $\ln~y(r, \Omega)$ and $\gamma(r, \Omega)$,

\begin{eqnarray}
c(r, \Omega) = \exp\left( - \beta v(r, \Omega) + \ln~y(r, \Omega) \right) - 1 - \gamma(r, \Omega)
\label{eq:cromega}
\end{eqnarray}

\section{Results}
\label{sec:results}

In the following, we will present the exact DCFs for the scCO$_2$ in the near-critical conditions. We will first evaluate the ability of HNC in reproducing the thermodynamic properties and the local structure in the near critical region. While the HNC short-range structures compare well with MD, their long-range correlations and thermodynamics differ --- especially near the critical point. We will then calculate the exact (MD-based) DCFs and the exact bridge functions of the EPM2 model and provide a comparison with the correlations obtained with HNC. Using the exact and HNC DCFs, we will finally calculate the chemical potentials of CO$_2$ with the MDFT-HRA functional and compare them with MD results.

\subsection{Thermodynamics and structure}
\label{sec:thermo}

\begin{figure}
     \begin{subfigure}{0.3\textwidth}
     \includegraphics[width=1.1\textwidth]{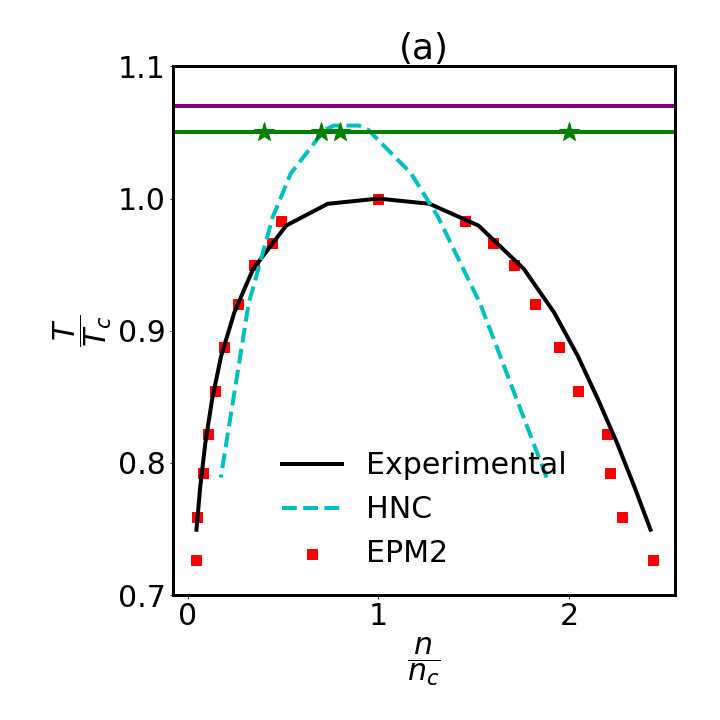}    
         \label{fig:phase_diag}
     \end{subfigure}
    \begin{subfigure}{0.3\textwidth}
         \includegraphics[width=1.1\textwidth]{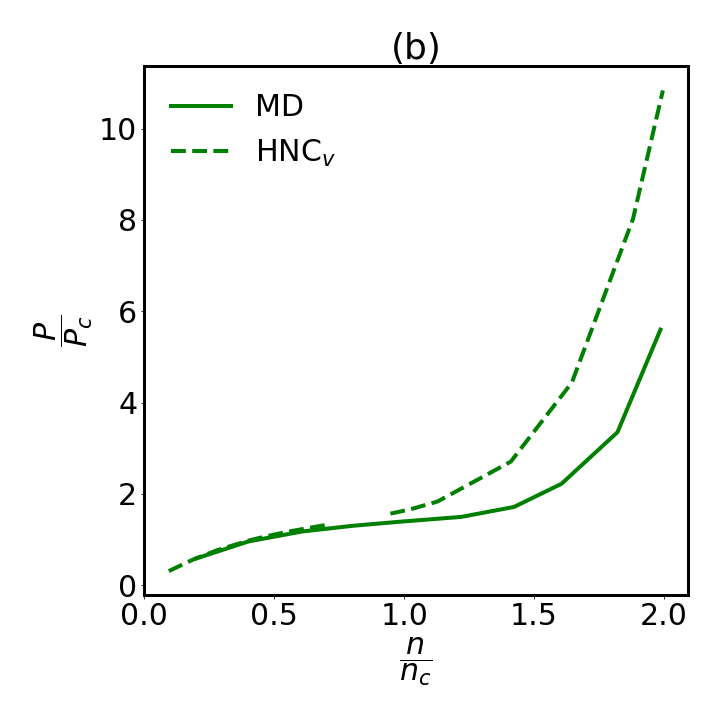}    
         \label{fig:T_320K}
     \end{subfigure}
     \begin{subfigure}{0.3\textwidth}
         \includegraphics[width=1.1\textwidth]{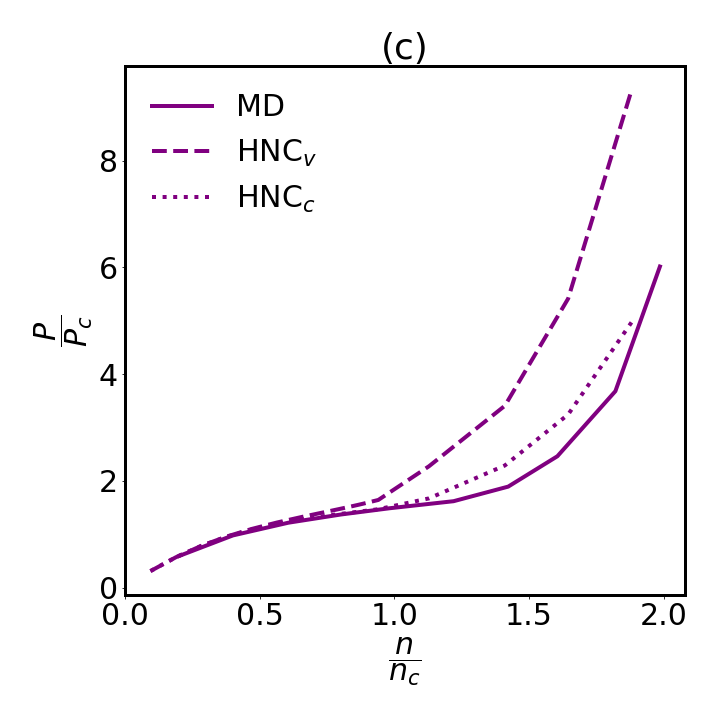}
         \label{fig:T_325K}
     \end{subfigure}
\caption{Phase diagram of scCO$_2$ obtained with MD and HNC. (a) Vapor-liquid coexistence curve from the experiments, from the MD simulations and from the HNC calculations. Note that for HNC, the dashed line indicates the limit of existence of HNC solutions (not a physical transition line). The horizontal lines are the two isotherms presented in (b) and (c). The star symbols indicate the four densities investigated in this work. (b) Isothermal evolution of the pressure vs. the density for MD and HNC (with the viral route) at $T = 1.05~T_c$. (c) Same as (b), for $T = 1.07~T_c$. For HNC, both the viral and the compressibility routes are presented (see text for further details).}
\label{fig:phase_diagram}
\end{figure}

The EPM2 force field was specifically designed to accurately reproduce the experimental liquid-vapor coexistence curve~\cite{harris_carbon_1995}. A comparison between the experimental and MD curves is shown in Fig.~\ref{fig:phase_diagram}(a). The HNC forbidden region is added for comparison, although its boundary is more or less identified with the spinodal line.~\cite{belloni_inability_1993, rull_absence_1996} Since $T_c(\mathrm{HNC})>T>T_c$, the HNC solution misses a small density part of the isotherm, in the close proximity of the true critical point. \par


Outside the HNC-forbidden domain, we have evaluated the ability of HNC to reproduce the MD-based equation of state (EOS). HNC pressures can be calculated using two approaches: (i) the virial route (see ref.~\cite{belloni_finite-size_2017} for an optimized use of the projections in practice for the virial pressure), and (ii) the compressibility route, which proceeds by integrating the compressibility along an isotherm.~\cite{hansen_theory_2013} The compressibility route requires that HNC can be solved for all densities along the isotherm, in particular the isotherm should not cross the HNC-forbidden domain. We consider two isotherms, one which crosses the HNC-forbidden domain ($T = 1.05~T_c$, green line in Fig.~\ref{fig:phase_diagram}(a)), and one which stays above ($T = 1.07~T_c$, purple line in Fig.~\ref{fig:phase_diagram}(a)). Fig.~\ref{fig:phase_diagram}(b) and (c) show the (reduced) pressure vs. the (reduced) density for the two isotherms $T = 1.05~T_c$ and $T = 1.07~T_c$, respectively. For $T = 1.05~T_c$, only HNC-virial pressures can be calculated. The HNC-virial pressures coincide with the MD pressures in the gas-like region (for lower densities than the HNC-forbidden region), but diverge from the MD pressures in the liquid-like region (for higher densities than the HNC-forbidden region). Both HNC-virial and HNC-compressibility pressures can be calculated in the case of  $T = 1.07~T_c$. The HNC-virial follows the same trend as for $T = 1.05~T_c$, being close to the MD pressures in the gas-like region and a divergence in the liquid-like region. On the other hand, the pressures computed from the compressibility route are close to the MD results, even in the liquid-like region. \par 

To understand the limitations of HNC in terms of structures and spatial correlations, we decided to investigate the isotherm $T = 1.05~T_c$ more in depth. In the following, we will consider three densities outside the HNC-forbidden domain (0.4, 0.7 and 2.0~$n_c$) and one within (0.8~$n_c$), at $T = 1.05~T_c$.  \par

\begin{figure*}
\begin{center}
    \includegraphics[width=1\textwidth]{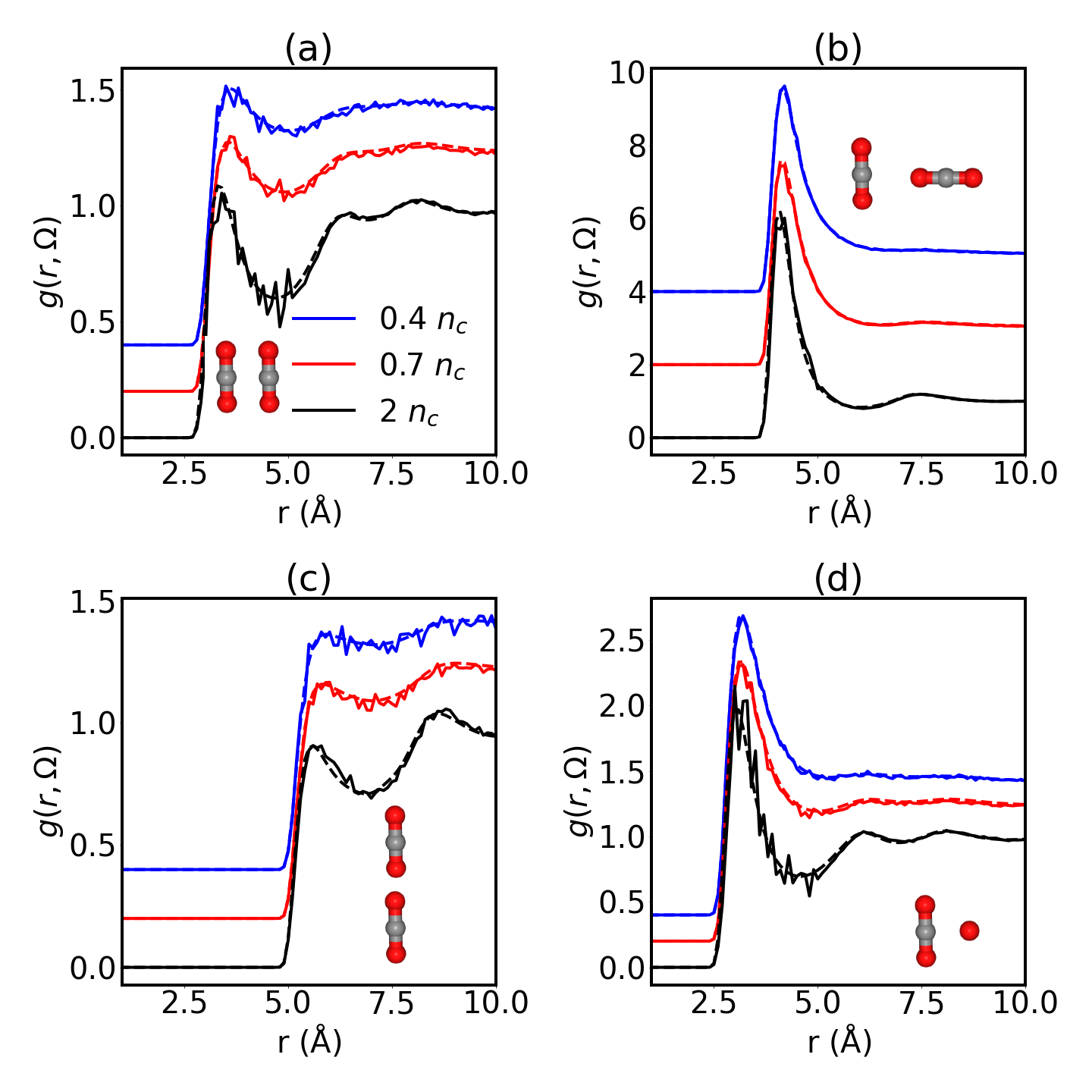}
\end{center}
\caption{Pair distribution functions (PDF) $g(r, \Omega)$ for four characteristic orientations $\Omega$, displayed as insets. MD results are indicated as solid lines and HNC results as dashed lines. Results are shown for three densities where HNC solutions exist (0.4, 0.7 and 2.0~$n_c$). No results are shown for $0.8~n_c$, as MD PDF are indistinguishable from $0.7~n_c$ and the HNC calculation fails. The PDFs are shifted upwards for easier comparison.}
\label{fig:rdf}
\end{figure*}

We first focus on the short-range parts of the PDFs and compare the MD and the HNC results. The MD PDFs are obtained by solving the OZ with the mixed closure Eq.~\ref{eq:mixed_closure}, extracting the cavity potentials $[\ln~y]_\alpha$ (as explained in Sec.~\ref{sec:extraction_short}, summing them to get $\ln~y(r, \Omega)$, and finally calculating the PDFs with Eq.~\ref{eq:gromega}). 
We report as Supporting Information (Fig.~S1 and Fig.~S2) the confirmation of the convergence of the mixed closure algorithm, and the $\ln~y(r, \Omega)$. On ther other hand, the HNC PDFs are calculated by self-consistently solving the HNC equations (see Section~\ref{sec:hnc_resol}). We have chosen to present the PDFs for four illustrative orientations $\Omega$ between two CO$_2$ molecules: parallel, perpendicular on the same plane (T-shaped), aligned, and along two perpendicular axes (in a cross like shape). Fig.~\ref{fig:rdf} presents the PDFs computed from MD (solid line) and from HNC (dashed line) for the three densities for which HNC admits solutions and for the four orientations (shown in the insets). We clearly observe a good agreement between the HNC and MD PDFs at short-range, as they coincide for all densities and orientations. \par

The orientational PDFs reveal a strong microscopic organization of the scCO$_2$ fluid:  the first peak of the PDFs is around 4~\AA~ and is the highest for the perpendicular (in plane) orientation ($g \sim 6$, see Fig.~\ref{fig:cr_4proj}(b)), slightly lower for the cross orientation ($g \sim 2$, see Fig.~\ref{fig:cr_4proj}(d)) and significantly lower for the two other orientations (parallel and aligned, $g \sim 6$, see Fig.\ref{fig:cr_4proj}(a) and (c)). This shows that two neighboring CO$_2$ molecules preferentially arrange themselves in the perpendicular (in plane) orientation, a configuration that minimizes the quadrupolar-quadrupolar interaction. \par

\begin{figure*}
    \includegraphics[width=1\textwidth]{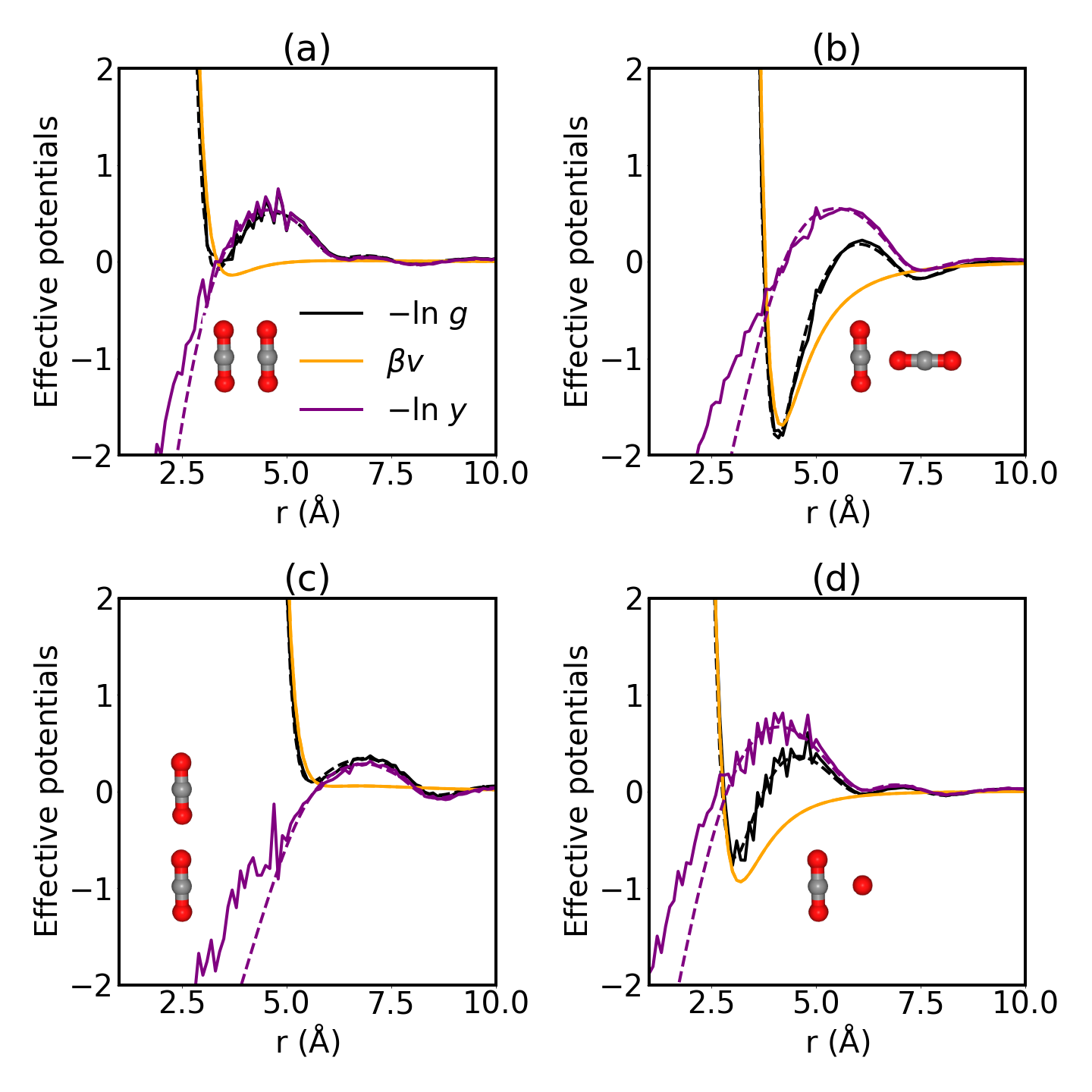}
\caption{Decomposition of the effective potentials in terms of the ideal and of the excess contribution. MD results are indicated as solid lines and HNC results as dashed lines. The density is 2.0~$n_c$ and the four mutual orientations $\Omega$ are displayed as insets. The decomposition reads $ - \ln g(r, \Omega) = \beta v - \ln y(r, \Omega)$. }
\label{fig:decomp}
\end{figure*}

To investigate the short-range organization of the scCO$_2$, we decompose the PDF $g(12)$ into two contributions: the direct interaction $\beta v(12)$ and the cavity potential $- \ln y(12)$, which accounts for the effective many-body correlations. Fig.~\ref{fig:decomp} shows the decomposition of the effective potential $-\ln~g=\beta v - \ln~y$ for the density 2.0~$n_c$ and for the four different orientations. The pair potential is analytically known and displays the expected behavior for quadrupolar-quadrupolar interactions: for all orientations, it diverges positively at small distances and goes to zero at long distances (above 8~\AA). At intermediate distances (4-8~\AA), the pair potentials for the perpendicular, cross and parallel orientations reach a minimum, with a depth of 1.8, 0.9 and 0.1~$k_B T$, respectively, while no minimum is observed for the aligned orientations. 
On the other hand, the cavity potentials evolve similarly for all orientations: they smoothly increase from a negative initial value to a peak around 5~\AA~and then they decrease to zero with small overdamped oscillations. At short distances and up to 4~\AA, the effective potentials $- \ln g(12)$ are governed by the pair potentials $\beta v(12)$, which explains the usual short-range features of the PDFs: the zero values at very short distance and the position and amplitude of the first peak. At intermediate distances, the cavity potential plays a role. Adding up the pair and the cavity potentials, we observe an energy barrier at 5~\AA~with a height of about 0.5~$k_B T$, followed by overdamped oscillations. The barrier and the oscillations indicate the presence of a first and subsequent solvation shells around a CO$_2$ molecule. Our analysis shows that this structuring arises mainly from the many-body correlations, $- \ln y(12)$, while the differences between the PDFs of the four orientations arise principally from the direct interaction $v(12)$. \par

\begin{figure*}
    \includegraphics[width=1\textwidth]{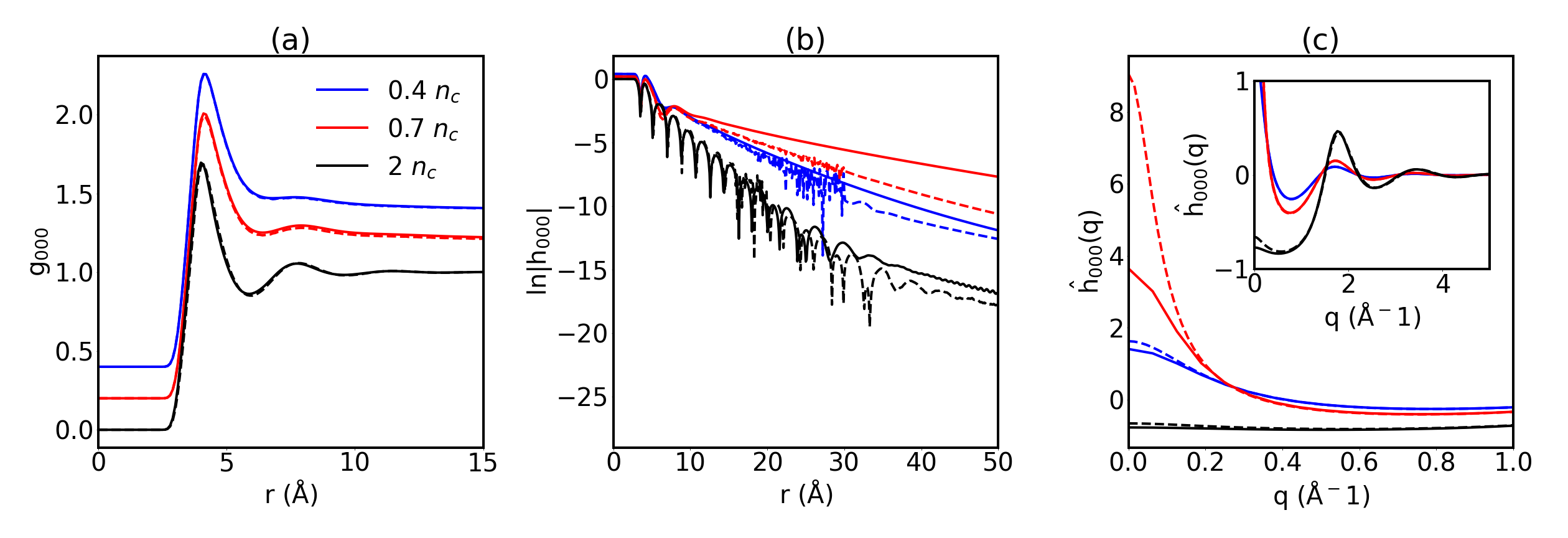}
\caption{(a) Radial distribution functions $g^{000}(r)$ obtained with MD (solid line) and HNC (dashed lines) for the 3 densities (0.4, 0.7 and 2.0~$n_c$). The PDFs are shifted upwards to facilitate the comparison. (b) Absolute value of the logarithm of the indirect correlation function, $h = g - 1$. (c) Fourier-transform of the indirect correlation function $\hat{h}(q)$. Inset: Zoom out to highlight the structure at $2.0~n_c$. While almost identical at short distances, MD and HNC differ at long-range. }
\label{fig:gcc_hcc}
\end{figure*}

The excellent agreement at short distances should not hide the difficulty of HNC in obtaining the correct thermodynamic properties, as the phase diagram (Fig.~\ref{fig:phase_diagram}). The thermodynamic properties are directly related to the long-range part of the center-of-mass indirect correlation function, $h^{000} = g^{000} - 1$. The center-of-mass PDFs are reported in Fig.~\ref{fig:gcc_hcc}(a) for the three densities (0.4, 0.7 and 2.0~$n_c$). As for the orientational PDFs, MD and HNC results coincide in the range 0-15~\AA. However, they differ largely at long distance. We highlight their difference by considering the $\ln \vert h^{000} \vert$ and its Fourier-transform, $\hat{h}_{000}(q)$, shown in Fig.~\ref{fig:gcc_hcc}(b) and (c), respectively. The difference is more pronounced in the near-critical region (0.7~$n_c$) and less pronounced in the liquid-like region (2.0~$n_c$). The $\ln \vert h^{000} \vert$ also shows that HNC correctly reproduces the oscillations of $h^{000}$ at 2.0~$n_c$, which are markers of the long-range organization in liquids. The difference between MD and HNC in the near-critical region is also visible when $\hat{h}^{000}(q)$ is considered (see Fig.~\ref{fig:gcc_hcc}). For $q > 0.3$~\AA$^{-1}$, no difference is visible between the MD and the HNC results. However, below $0.3$~\AA$^{-1}$, discrepancies appear, although they remain small for 0.4~$n_c$ and for 2.0~$n_c$ (see the inset of Fig.~\ref{fig:gcc_hcc}). In the $q=0$ limit, the indirect correlation function relates to a key thermodynamic property, the reduced compressibility, $\chi_T = 1 + \rho_b \hat{h}^{000}(q=0)$, which are reported in Table~S2. The difference between HNC and MD $\hat{h}_{000}$ in the low-$q$ region and at $q=0$ confirms that HNC cannot reproduce the thermodynamic properties in the near-critical region.   \par

To conclude this section, we show that the failure of HNC to deliver the correct thermodynamics stems from its incorrect prediction of the long-range structure, especially close to the critical point. Interestingly, the molecular HNC accurately captures the short-range organization of scCO$_2$ and its evolution from the gas-like to the liquid-like region. In the following, we will present the exact DCFs and bridge functions, and compare them with the HNC results. \par


\subsection{Direct correlation and bridge functions}
\label{sec:dcfbridge}

We will now analyze the direct correlation functions (DCFs), which are the main ingredients to perform MDFT calculation at the HRA level. We will compare the exact DCFs obtained from MD with those calculated with HNC. Eventually, we will present the exact bridge functions $b$, which are the neglected contribution in the HNC approximation. \par

\begin{figure*}
    \includegraphics[width=\textwidth]{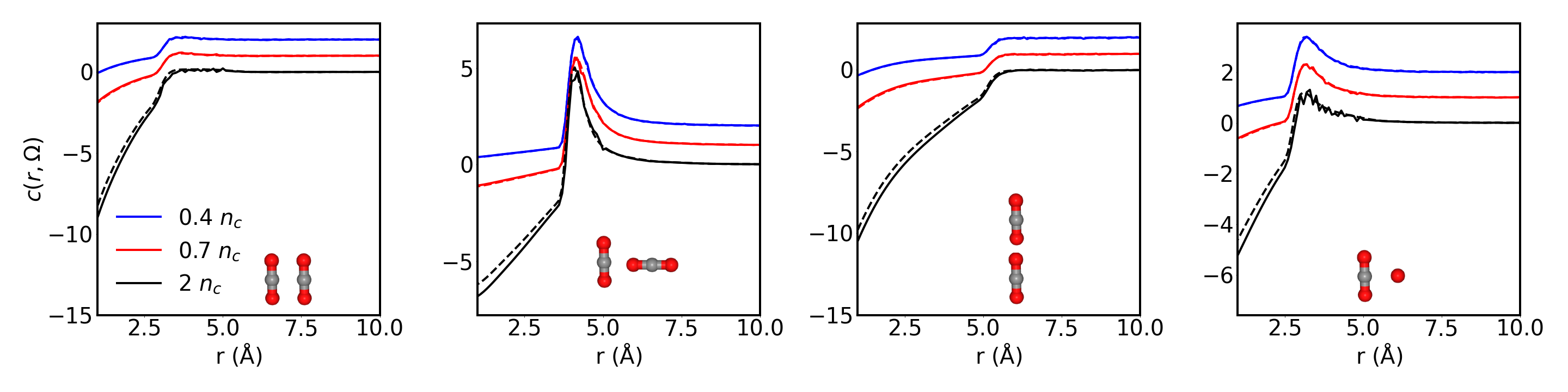}
    \includegraphics[width=\textwidth]{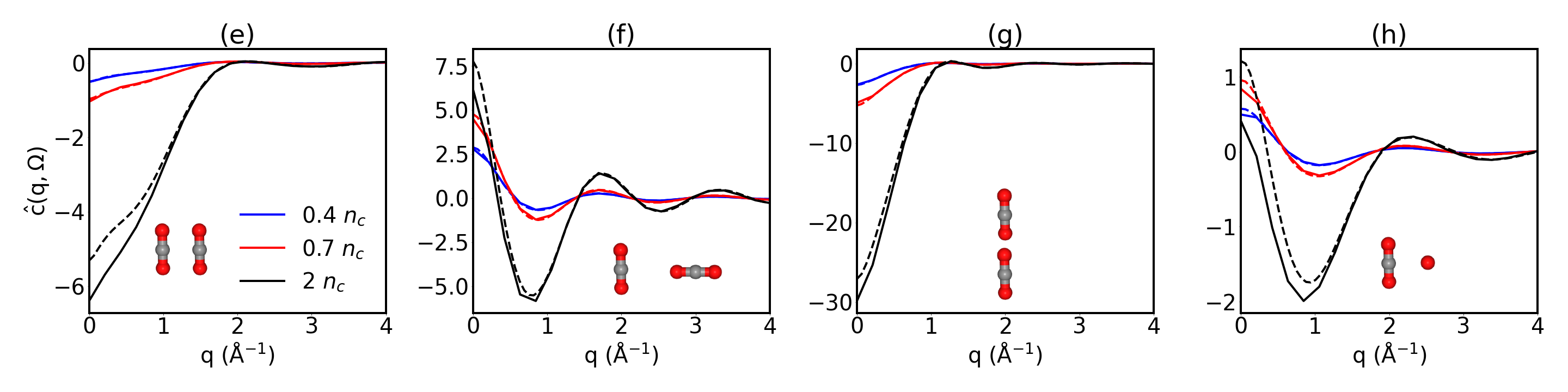}
\caption{Molecular direct correlation function $c(r, \Omega)$ for four characteristic orientations $\Omega$ shown as insets. As in Fig.~\ref{fig:rdf}, the MD results are indicated using full lines and the HNC results using dashed lines. The results for three densities are shown (0.4, 0.7 and 2.0~$n_c$), with an upward shift.}.
\label{fig:cr_4proj}
\end{figure*}

The DCFs $c(r, \Omega)$ are obtained from the recombination of the $\ln~y(r, \Omega)$ and $\gamma(r, \Omega)$ from the projections $[\ln~y]_\alpha$ and $\gamma_\alpha$. Fig.~\ref{fig:cr_4proj} shows the orientational DCFs for the three densities for which HNC admits a solution and for the same four CO$_2$/CO$_2$ mutual orientations described previously for the PDFs. Fig.~\ref{fig:cr_4proj}(a)-(d) shows the DCFs $c(r)$ in real space, while Fig.~\ref{fig:cr_4proj}(e)-(h) shows their Fourier transforms $\hat{c}(q)$. In the direct space, all DCFs behave similarly: an increase from negative values below 4~\AA, an abrupt change in slope between 3 and 5~\AA, a peak around 4-5~\AA~ and a smooth decrease to zero for higher distances. The highest peak is observed for the perpendicular orientation (\ref{fig:cr_4proj}(b)), a smaller one for the cross orientation (\ref{fig:cr_4proj}(d)) and it is almost invisible for the other two orientations (parallel and aligned, (a) and (c)). The different densities only impact the initial values of the DCFs, otherwise a similar trend compared with that of the PDFs (Fig.~\ref{fig:rdf}) was found. In the Fourier space (Fig.~\ref{fig:cr_4proj}(e)-(h)), the DCFs mainly differ in the case of the perpendicular and of the crossed orientations with respect to the case of the parallel and of the aligned orientations. First, the initial values $\hat{c}(q=0, \Omega)$ are positive for the perpendicular and crossed orientations, while they are negative for the parallel and aligned orientations. In addition, the perpendicular and cross orientations display marked oscillations, related to the peaks in the direct space, while the transformed DCFs for the parallel and aligned orientations are smoother. \par

If we now compare HNC and MD DCFs, we notice that they overlap perfectly for all distances at densities 0.4~$n_c$ and 0.7~$n_c$. For the density 2.0~$n_c$, they overlap only beyond 4~\AA (\textit{i.e.}, beyond the slope change). Below 4~\AA, the HNC DCFs evolve slightly above the MD ones, though the difference is small. In the Fourier space, the HNC and MD DCFs are completely identical for 0.4~$n_c$ and 0.7~$n_c$. For 2.0~$n_c$, HNC and MD DCFs are similar above 1.5~\AA$^{-1}$, while the HNC DCFs evolve slightly above the MD DCFs below 1.5~\AA$^{-1}$. Notably, at 2.0~$n_c$, the DCFs computed with HNC differ from the MD results both at short distances (below 4~\AA) and at long distances (below 1.5~\AA$^{-1}$). The intermediate structure is well reproduced by HNC in the liquid-like region. For the lower densities, HNC and MD DCFs are similar for all distances. \par

\begin{figure*}
 \includegraphics[width=1\textwidth]{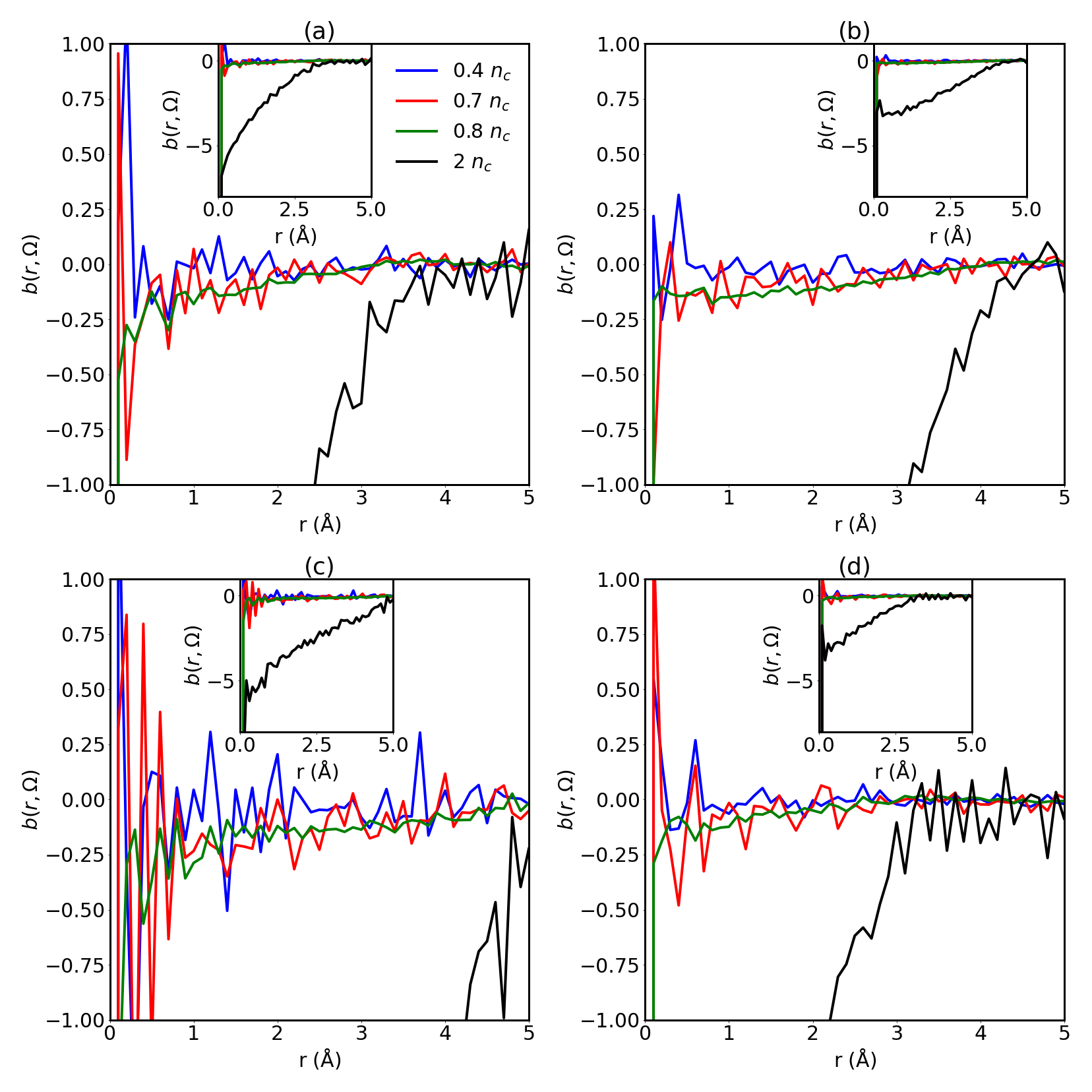}
\caption{Bridge functions $b(r, \Omega)$ obtained from MD simulations for the same orientations as in Fig.~\ref{fig:rdf}, and for the four densities (0.4, 0.7, 0.8 and 2.0~$n_c$). In HNC, the bridge functions are zero by definition. }
\label{fig:bridge}
\end{figure*}

We have finally analyzed the bridge functions $b(12)$. We calculated the 30 projections of the bridge from the $g$, $\ln~y$ and $c$ 
functions and recombined them for the four CO$_2$/CO$_2$ orientations chosen in this work. The bridge functions derived from MD data are presented in Fig.~\ref{fig:bridge} for four densities (now including 0.8~$n_c$), while the bridge functions of HNC are zero by definition. For the lowest density (0.4~$n_c$, blue line), the bridge functions display fluctuations around zero. In contrast, for the densities 0.7 and 0.8~$n_c$, they are negative and increase to zero in a few~\AA. For all these densities, the bridge functions are small compared to the cavity potential (Fig.~S2). This result corroborates the previous comparisons between the HNC and the MD correlation functions (PDFs, DCFs and $\ln y$): the bridge functions do not impact the short-range structures. They are however strictly non-zero and they affect the equation of state and the phase diagram as we discussed in Section~\ref{sec:thermo} when comparing MD and HNC. The most striking effect of the bridge is obtained for 0.8~$n_c$. The bridge functions at 0.8~$n_c$ are small (see Fig.~\ref{fig:bridge}, green lines), but HNC equations (\textit{i.e.}, neglecting the bridge) fail for this density and this isotherm. Indeed, even a small and short-ranged bridge affects the PDFs at all distances \textit{via} the OZ Eq.~\ref{eq:oz} and the $\gamma$ function in Eq.~\ref{eq:bridge}. On the other hand, the bridge functions are nonzero for the liquid-like density 2.0~$n_c$ as shown in the insets of Fig.~\ref{fig:bridge}. This also confirms the previous analysis of the PDFs and DCFs, which indicated that the MD and the HNC data diverge most signifcantly for this density. The bridge functions at 2.0~$n_c$ evolve smoothly from a negative value to zero. It is noteworthy that the bridge functions exhibit a similar behavior for the four different orientations, in contrast with the other correlation functions. This result suggests that the orientational dependence of the bridge function is presumably weak, which is an interesting finding for the development of integral equation theories capable of outperforming HNC. 

\subsection{Solvation free energy of CO$_2$}
\label{sec:solv}

We are now equipped to perform MDFT calculations at the HRA level for the 4 densities presented here. To assess the ability of the MDFT to provide the solvation properties, we calculated the solvation free energy of CO$_2$ into CO$_2$ (\textit{i.e.} the chemical potential) using 3 methods: MD, MDFT-HRA-MD and MDFT-HRA-HNC. MD calculations are done with the Widom insertion technique ~\cite{widom_topics_1963} (as outlined in  Sec.~\ref{sec:extraction_short}); MDFT-HRA-MD with the minimization of Eq.~\ref{eq:mdfthra} with $c = c^\mathrm{MD}$; MDFT-HRA-HNC with the minimization of Eq.~\ref{eq:mdfthra} with $c = c^\mathrm{HNC}$. Both the DCFs $c^\mathrm{MD}$ and $c^\mathrm{HNC}$ are visible in Fig.~\ref{fig:cr_4proj} and discussed in Sec.~\ref{sec:dcfbridge}, except for the density $n_b = 0.8~n_c$ for which the HNC calculation fails. \par

At the low densities ($< n_c$), the solvation free energies are remarkably similar for the three methods. This demonstrates the force of integral equations: to achieve the same results, MDFT-HRA requires 100,000 times fewer numerical resources than MD simulations. Interestingly, MDFT-HRA-HNC is equivalent to MDFT-HRA-MD for the densities $0.4$ and $0.7~n_c$. It confirms the similarity of the exact DCFs and the HNC DCFs that was previously discussed in Sec.~\ref{sec:dcfbridge}. \par

At the density $2.0~n_c$, the MD solvation free energy differs by more than 2.0~$n_c$ from the MDFT-HRA results. At high density/high pressure, the HRA is known to fail as it overestimates the free energy of cavitation $\Delta G_\mathrm{cav}$. For a macroscopic solute with a volume $V_\mathrm{sol}$, the cavitation free energy is $\Delta G_\mathrm{cav} = - P V_\mathrm{sol}$. However the effective pressure in HRA overestimates the real pressure ($P_\mathrm{HRA} > P$), which explains its failure at $2.0~n_c$. To improve HRA, the rigorous approach would be to incorporate a bridge functional $\mathcal{F}_B$, \textit{e.g.}, based for instance on hard-sphere functionals~\cite{levesque_scalar_2012, jeanmairet_molecular_2013-1} or on coarse-grained density functionals,~\cite{jeanmairet_molecular_2015, gageat_bridge_2017, borgis_simple_2020}. The development of a bridge functional is beyond the scope of the present work. We tried a simpler post-minimization correction, called the pressure correction~\cite{sergiievskyi_fast_2014, sergiievskyi_solvation_2015}, but this correction detoriates the comparison with MD and MDFT (as discussed in the supplementary information).

The success of MDFT in reproducing the solvation free energies at low densities of CO$_2$ are promising. MDFT with homogeneous reference approximation (MDFT-HRA) gives an excellent agreement with MD results, even if we used the HNC DCFs as input for the functional. At $2.0~n_c$, the MDFT-HRA fails to give the correct result in comparison with MD. In the future, we will develop an efficient bridge functional to surpass the HRA and to accurately determine the solvation free energy for all densities. 

\begin{table}[H]
\centering
\caption{Solvation free energies $\Delta G_\mathrm{sol}$ (in kJ.mol$^{-1}$) of a CO$_2$ molecule into scCO$_2$ at $T = 1.05~T_c$ for four densities and calculated with three methods: MD, MDFT-HRA-MD and MDFT-HRA-HNC. Details are provided in the text.}
\begin{tabular}{lcccc}
$n_b$ & 0.4~$n_{c}$ & 0.7~$n_{c}$ & 0.8~$n_{c}$ & 2~$n_{c}$  \\ 
\hline
\hline
MD & -1.83 & -2.91 & -3.21 & -3.79  \\
\hline
MDFT-HRA-MD & -1.83  & -2.88  & -3.16   & -0.918   \\ 
\hline
MDFT-HRA-HNC  &  -1.83  & -2.87  & -  & -1.50  \\ 
\end{tabular}
\label{tab:deltaE}
\end{table}

\section{Conclusion}
\label{sec:conclusion}

The supercritical CO$_2$ is a key solvent within the increasing development of green processes. Its non-toxic properties make it ideal for the extraction and impregnation processes in the food or pharmaceutical industry. However, the lack of predictive models for the solvation impedes a larger use of scCO$_2$. The existing approaches are too expensive (molecular simulations), not flexible (parameterized approach) or not accurate (liquid-state theories). \par

This work establishes the foundation for the construction of a molecular density functional theory (MDFT) of scCO$_2$. MDFT is a powerful method that combines the spatial and orientational solvent density with the classical DFT. It permits the precise modeling of the interactions at molecular scale --- particularly solute-solvent anisotropic interactions. Within the homogeneous reference approximation (HRA), the MDFT requires two inputs: the potential of solute-solvent interactions and the direct correlation functions (DCFs), which stand for the effective solvent-solvent interactions.  \par

We present here the exact DCFs of scCO$_2$ for four densities in the near-critical region. The aforementioned quantities were extracted by combining pair distribution functions derived from MD simulations of bulk CO$_2$ with the molecular Ornstein-Zernike equation and the mixed closure HNC at long distances. This technique is now becoming a routine procedure, though it requires long MD trajectories as well as special attention for the short distances of the distribution functions. In addition to the DCFs, we also calculated and discussed other correlation functions, namely the pair distribution functions and the cavity potentials $\ln y$. The DCFs, PDFs and $\ln y$ exhibit analogous patterns: (i) a higher amplitude for the perpendicular and cross configurations of a pair of CO$_2$ molecules (which correspond to the optimal configurations for the quadrupolar-quadrupolar interaction) and (ii) a stronger microscopic structuring for the highest density (2.0~$n_c$, in the liquid-like region) compared to the smaller densities ($< n_c$, in the gas-like region). \par

We also compare the exact DCFs, PDFs and $\ln y$ with those predicted by the molecular HNC theory, the simplest integral equation theory for a molecular system. As expected, the molecular HNC failed to reproduce the equation of state or the phase diagram of the scCO$_2$ (modeled with the EPM2 force field). The molecular HNC is successful in obtaining the short-range structure for all orientations, as well as the smooth transition from a gas-like, unstructured fluid, to a more liquid-like fluid at higher density. We also investigate the bridge functions, which are the neglected contribution in HNC. They are weak for low densities and are orientation-independent for the highest density. \par

Finally, we applied MDFT to calculate the solvation free energy of CO$_2$ in scCO$_2$. We use MDFT at the HRA level with the exact DCFs and with the HNC DCFs, and we compared their results with the exact MD-based calculations. The comparison demonstrates the ability of MDFT-HRA to determine the solvation free energy in the gas-phase region, for both HNC and MD DCFs. At higher density, the MDFT-HRA fails to compute the correct solvation free energy. The comparison underlines the necessity to develop a bridge functional working at all densities, and highlights that the HRA functional based on the HCN DCFs performs as well as the HRA functional based on the exact DCFs. \par

Work is currently ongoing to develop the bridge functional and to evaluate MDFT for a larger number of molecular solutes. Our long-term objective is to develop a tool to determinate the excess functional for each density, each temperature and each environment (\textit{e.g.} presence of cosolvent, confinement). While exact, the extraction of DCFs from MD data remains too expensive to participate in the development of the scCO$_2$ as a green solvent. Our current goal is to develop a new method for determining efficiently the excess solvent functional of MDFT at any pressure and temperatures. The combination of  molecular integral equations such HNC (which provides an accurate representation of the microscopic structure) and of a weighted-density approximation for the functional bridge appears to be a promising route to pursue.

\clearpage

\bibliography{biblio}
\bibliographystyle{elsarticle-num}




\end{document}


\title[]{\textbf{Supplementary information} of ``Molecular integral equations theory in the near critical region of CO$_{2}$''}

\author{M. Houssein Mohamed}
\affiliation{LPCT, Université de Lorraine, CNRS, Nancy 54000, France}
\author{L. Belloni}
\affiliation{LIONS, NIMBE, CEA, CNRS, Université Paris-Saclay, Gif-sur-Yvette 91191, France}
\author{D. Borgis}
\affiliation{PASTEUR, Département de Chimie, École Normale Supérieure, PSL University, Sorbonne Université, CNRS, Paris 75005, France}
\author{F. Ingrosso}
\author{A. Carof}
\email{antoine.carof@univ-lorraine.fr.}
\affiliation{LPCT, Université de Lorraine, CNRS, Nancy 54000, France}

\date{\today}

\maketitle 

\section{Force field parameters}

\begin{table}[H]
\caption{Potentiel Function Parameters for the EPML2 force field}
\begin{ruledtabular}
\begin{tabular}{cccc}
$\epsilon_\mathrm{CC}$ & 28.129~K & $\sigma_\mathrm{CC}$ & 2.757~\AA  \\ 
$\epsilon_\mathrm{CO}$ & 47.588~K & $\sigma_\mathrm{CO}$ & 2.892~\AA  \\ 
$\epsilon_\mathrm{OO}$ & 80.507~K & $\sigma_\mathrm{OO}$ & 3.033~\AA  \\ 
$ l_\mathrm{CO}$ & 1.149~\AA & $ q_\mathrm{C}$ & +0.6512~e  \\ 
\end{tabular}
\label{tab:epm2}
\end{ruledtabular}
\end{table}

\section{Convergence of MD/HNC procedure}
\label{secapp:convergence}

\begin{figure*}[!htb]
\includegraphics[width=1\textwidth]{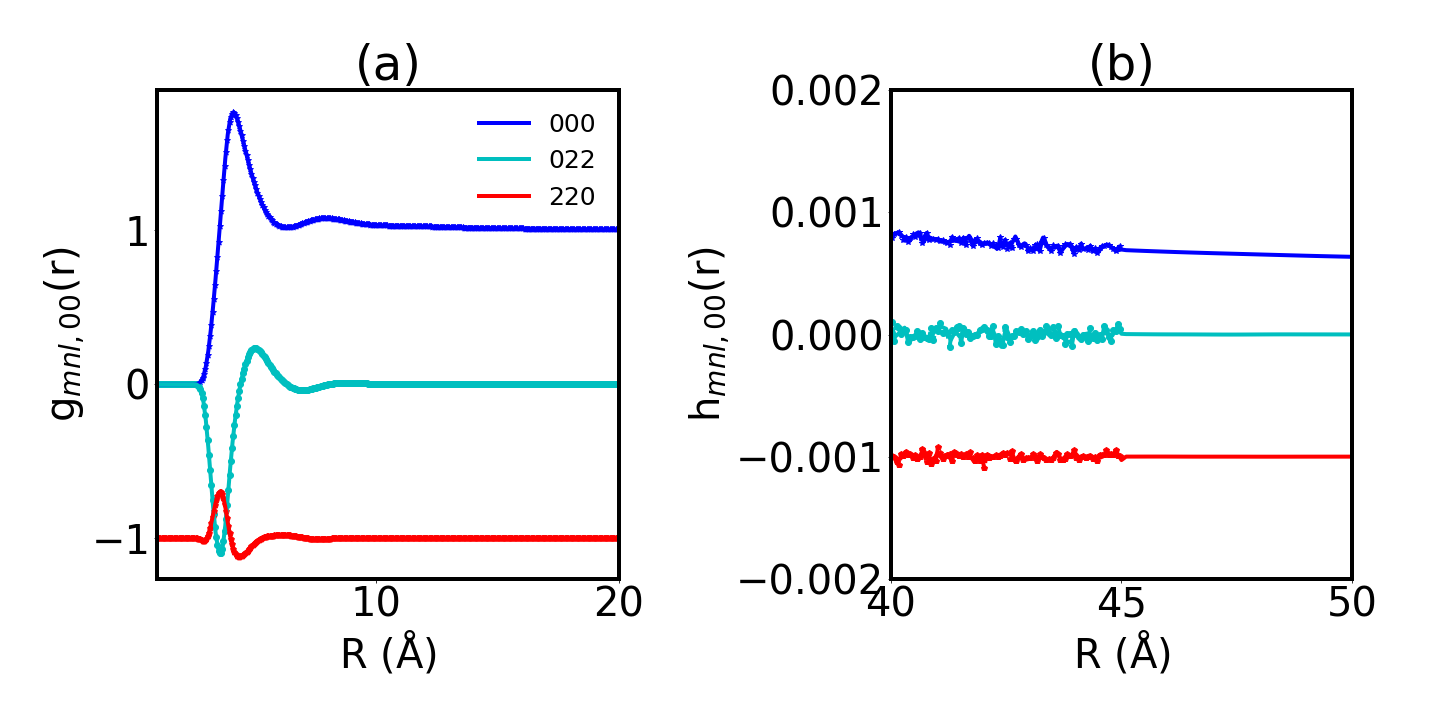}
\caption {(a) Projections g$^{mnl}_{00}$(r) for $n_b = 0.8~n_{c}$ of the PDFs. We have 30 independent projections but here we present only three of them. Symbols = MD data, shown up to $L_\mathrm{box}/2 = 45~$\AA; solid lines = MD integral equation, performed with $r_\mathrm{max}=L_\mathrm{box}/2$. (b) Focus on the intermediate distances.}
\label{fig:prolongation}
\end{figure*}

Fig.~\ref{fig:prolongation} confirms the success of the mixed closure resolution: at $r_\mathrm{max} = 45$~\AA, no difference is visible between pure MD (symbol) and MD completed with the mixed closure integral equation (solid lines).

\clearpage
\section{Cavity potentials}
\label{sec:correlation}

\begin{figure*}
    \includegraphics[width=0.8\textwidth]{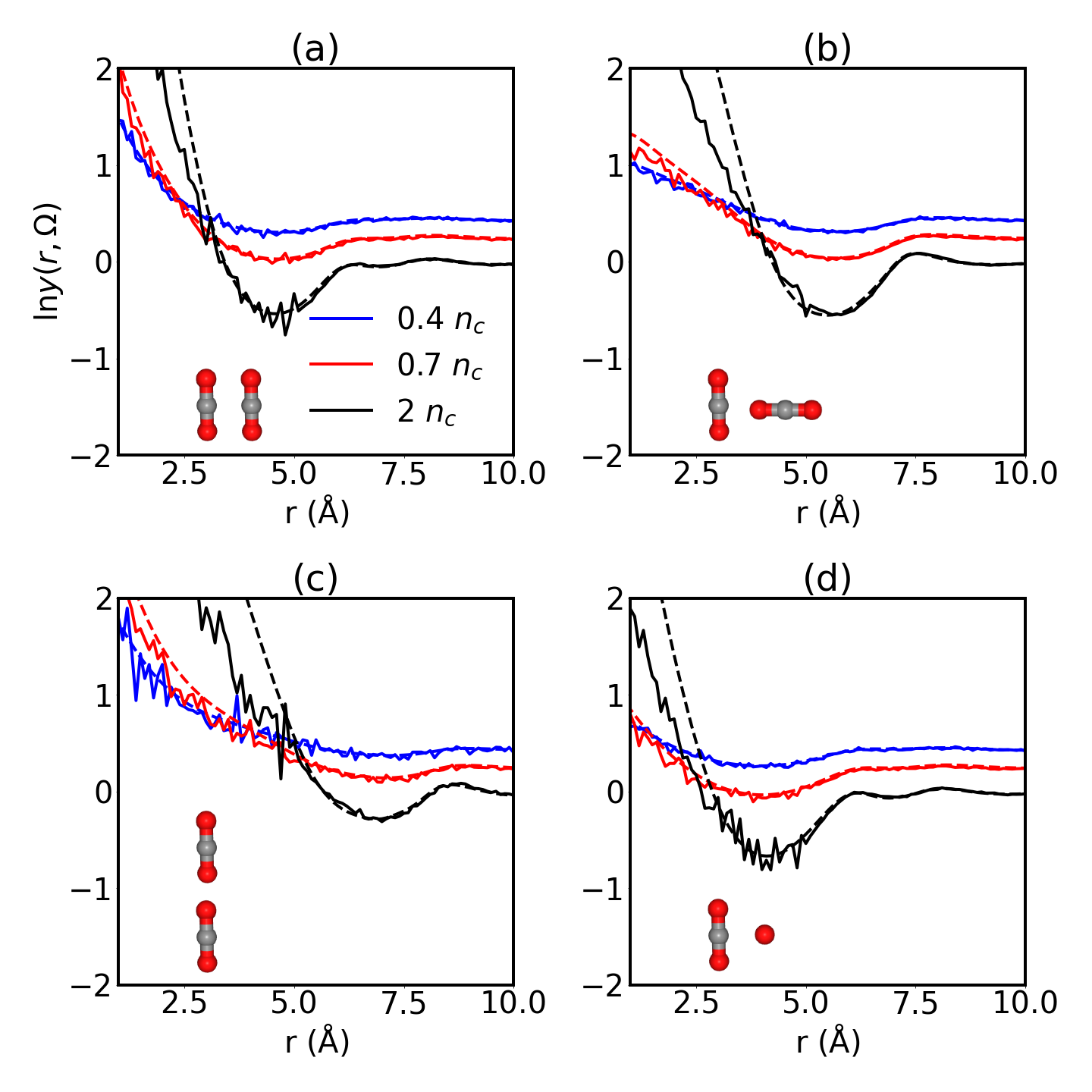}
\caption{Molecular $\ln y\left(r,\Omega\right)$ with $y$, the cavity function, for 4 characteristic orientations $\Omega$ pictured on the figure.  Results are obtained for 3 densities (0.4, 0.7 and 2.0~$n_c$) and are shifted upward for comparison. The exact MD approaches is indicated in continuous line and the HNC calculation with dashed lines.}
\label{fig:lny_orient}
\end{figure*}

The cavity potentials $\ln y(12)$ represents the non-ideal, collective part of the effective potentials. At medium and long distances (above 4~\AA), we calculated $\ln y$ from the PDFs and the intermolecular potential. However, at short distances, the values computed for the PDFs are equal to zero, and we applied the Henderson technique to extract the short-range projections of $\ln y$ (see Section~III.E). Fig.~\ref{fig:lny_orient} shows the $\ln y$ functions for the densities 0.4, 0.7, and 2.0 $n_c$ and for the same four orientations analyzed in the main text (and displayed in the insets of Fig.~\ref{fig:lny_orient}). Similarly to the PDFs and the DCFs, we found that $\ln y$ is more structured for the highest density, and the distinction between MD and HNC results only appear at short distances. \par

\clearpage
\section{Structure and compressibility}
\label{secapp:structure}

From the center-of-mass radial distribution function ($g_{000}(r) = g_{CC}(r)$), we calculated the coordination number, 

\begin{equation}
    n_\mathrm{coord} = \rho_b \int_0^{R_S} \mathrm{d}\mathbf{r}  g_{000}(r)
\label{eq:nc}
\end{equation}

\noindent
with $R_S$ the first minimum of the CC RDF. We also calculated the reduced compressibility, 

\begin{equation}
\chi_T = 1 + \rho_b \int \mathrm{d}\mathbf{r} h_{000}(r)
\label{eq:chit}
\end{equation}

\noindent
with $h_{000} = g_{000} - 1 = g_{CC} - 1$ the indirect correlation functions, and $\hat{h}_{000}(0) = \int \mathrm{d}\mathbf{r} h_{000}(r)$ is called the Kirkwood-Buff integral. Table~\ref{tab:nc_chi} gathers the coordination number and the reduced compressibility values for the four densities and for the HNC and MD calculations.

\begin{table}[H]
\caption{Coordination number $n_c$ (see Eq.~\ref{eq:nc}) and reduced compressibility $\chi_T$ (see Eq.~\ref{eq:chit}) for the four densities. $n_\mathrm{coord}$ and $\chi_T$ are dimensionless quantities }
\begin{ruledtabular}
\begin{tabular}{lcccc}
$n_b$ & 0.4$~n_c$ & 0.7$~n_c$ & 0.8$~n_c$ & 2$~n_c$  \\ 
\hline
$n_\mathrm{coord} $ (MD)    & 3  & 5  & 6 & 10   \\
\hline
$n_\mathrm{coord} $ (HNC)   & 3 & 5 & - & 10  \\
\hline
$\hat{h}_{000}(0)$ (MD)  (\AA$^3$) & 551.8  & 816.2  & 932.76  & -60.33   \\
\hline
$\hat{h}_{000}(0)$ (HNC) (\AA$^3$)  & 636.9  & 2025  & -  & -51.58   \\
\hline
$\chi_T$  (MD)   & 2.41  & 4.65  & 5.77 & 0.23  \\
\hline
$\chi_T$  (HNC)   & 2.63 & 10.6 & - & 0.34  
\end{tabular}
\label{tab:nc_chi}
\end{ruledtabular}
\end{table}

\begin{figure}
    \begin{subfigure}{0.45\textwidth}
         \caption{}
         \includegraphics[width=1\textwidth]{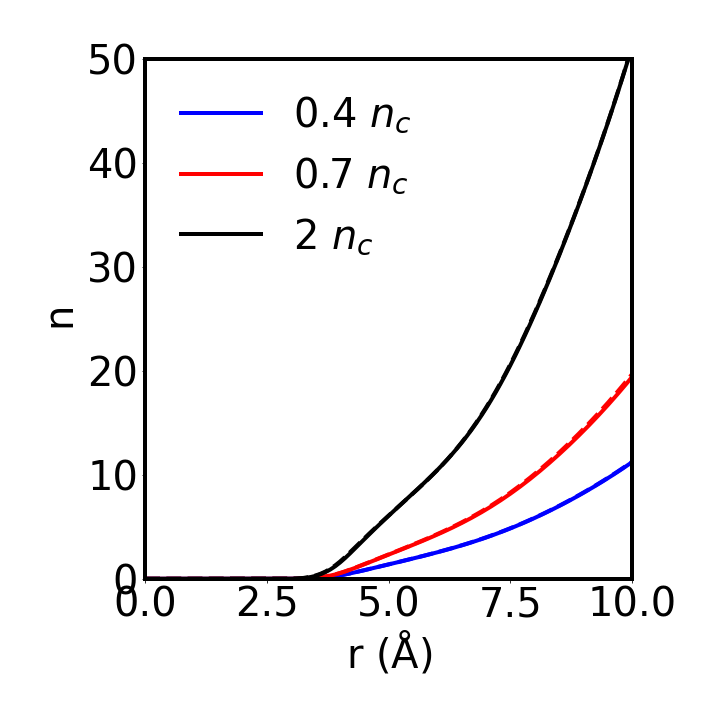}
     \end{subfigure}
     \hfill
     \begin{subfigure}{0.45\textwidth}
         \caption{}
         \includegraphics[width=1\textwidth]{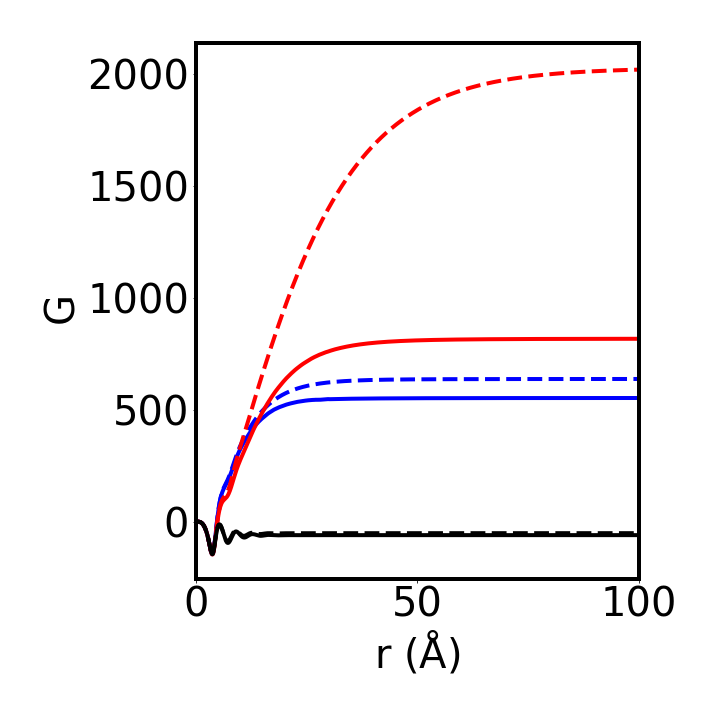}
     \end{subfigure} 
\caption{Comparison of (a) the running coordination number $n(r)$ and (b) the running Kirkwood-Buff integral obtained by MD and HNC. Solid lines: MD results and dashed lines: HNC results}
\label{fig:N_and_KBI}
\end{figure}

Fig.~\ref{fig:N_and_KBI} presents the running coordination number, $n(r) = n_b \int_0^r \mathrm{d}\mathbf{r}'  g_{000}(r')$ and the running Kirkwood-Buff integral $G(r) = \int_0^r \mathrm{d}\mathbf{r}' h_{000}(r')$. The running Kirkwood-Buff integrals converge at long distances (20, 25 and 80~\AA~for the densities 0.4, 0.7, and 2.0 $n_c$).

\clearpage

\section{Pressure correction}

To improve MDFT at the HRA level at high density, we tested the so-called pressure correction~\cite{sergiievskyi_fast_2014, sergiievskyi_solvation_2015}. After the minimization of the functional, we calculated and added the following free energy correction,

\begin{equation}
\Delta G^\mathrm{PC}_\mathrm{corr} = - (P_\mathrm{MD} - P_\mathrm{HRA}) V_\mathrm{sol}
\label{eq:pc_corr}
\end{equation}

\noindent where $V_\mathrm{sol}$ is the volume of the solute. It is derived by considering a macroscopic solute of volume $V_\mathrm{sol}$, whose cavitation free energy is $ \Delta G_\mathrm{cav} -PV_\mathrm{sol}$. Instead, MDFT at the HRA level will give: $\Delta G_\mathrm{cav} =  - P_\mathrm{HRA} V_\mathrm{sol}$, where the effective HRA pressure reads, 

\begin{align}
P_\mathrm{HRA} &= k_BT n_b\left(1 - \frac{1}{2}\rho_b \hat{c}_{000}(0) \right) \\
&= \frac{k_BTn_b}{2} \left( 1 + \frac{1}{1+\rho_b \hat{h}_{000}(0)} \right)
\label{eq:phra}
\end{align}

\noindent where the Ornstein-Zernike equation (Eq.~7 of the main text) was use between the first and second lines. Table~\ref{tab:pc} reports the values of the effective pressure of HRA (Eq.~\ref{eq:phra}) and of the pressure correction (Eq.~\ref{eq:pc_corr} of the main text), both with the exact DCFs ($c^\mathrm{MD}$) and the HNC DCFs ($c^\mathrm{HNC}$). We estimate $V_\mathrm{sol}$ as $V_\mathrm{sol} = - \int\mathrm{d}\mathbf{r}\mathrm{d}\Omega\left(\Delta\rho_\mathrm{eq}(\mathbf{r, \Omega}\right)/\rho_0 \approx -  \int\mathrm{d}\mathbf{r}h_{000}(\mathbf{r}) = -\hat{h}_{000}(0)$.  Table~\ref{tab:nc_chi} reports the values of $\hat{h}_{000}(0)$, while  Table~\ref{tab:pc} reports the correction, $\Delta G^\mathrm{PC}_\mathrm{corr}$ and the ``corrected'' solvation free energies, $\Delta G^\mathrm{HRA} + \Delta G^\mathrm{PC}_\mathrm{corr}$. The pressure correction does not improves the MDFT solvation free energies at low and high densities. To overcome HRA, we must develop an accurate bridge functional. \par

\begin{table}[H]
\centering
\caption{Effective HRA pressure (in bar, Eq.~\ref{eq:phra}) and pressure correction (in kJ.mol$^{-1}$, Eq.~25 of the main text) of a CO$_2$ molecule into scCO$_2$ at $T = 1.05~T_c$ for four densities.}
\begin{ruledtabular}
\begin{tabular}{lcccc}
$n_b$ & 0.4~$n_{c}$ & 0.7~$n_{c}$ & 0.8~$n_{c}$ & 2~$n_{c}$  \\ 
\hline
\hline
$P_\mathrm{HRA}$ (MD) (bar) & 79.89   & 120.1   & 133   &  1516  \\
\hline
$P_\mathrm{HRA}$ (HNC) (bar)& 77.95    & 108.6   & - &  1111  \\
\hline
$\Delta G_\mathrm{corr}$(MD) (kJ.mol$^{-1}$) & -0.14 & -1.14 & -1.70 & 3.9 \\
\hline
$\Delta G_\mathrm{corr}$(HNC)  (kJ.mol$^{-1}$) & -0.09 & -1.42 & - & 2.08 \\
\hline
$\Delta G^\mathrm{HRA} + \Delta G^\mathrm{PC}_\mathrm{corr}$(MD) (kJ.mol$^{-1}$) & -1.97 & -4.02 & -4.86 & 2.98 \\
\hline
$\Delta G^\mathrm{HRA} + \Delta G^\mathrm{PC}_\mathrm{corr}$(HNC)  (kJ.mol$^{-1}$) & -1.92 & -4.29 & - & 0.58 
\end{tabular}
\label{tab:pc}
\end{ruledtabular}
\end{table}

\bibliography{biblio.bib}